\documentclass[12pt,preprint]{aastex}
\shorttitle{Orion}
\shortauthors{McGehee}

\begin{document}

\title{Photometric Accretion Signatures Near the Substellar Boundary}


\author{Peregrine M. McGehee\altaffilmark{\ref{LANL1},\ref{NMSU}},
Andrew A. West\altaffilmark{\ref{UW}},
J. Allyn Smith\altaffilmark{\ref{LANL2},\ref{Wyo}},
Kurt S. J. Anderson\altaffilmark{\ref{NMSU},\ref{APO}}, \\
J. Brinkmann\altaffilmark{\ref{APO}}}

\altaffiltext{1}{
Los Alamos National Laboratory, LANSCE-8,
MS H820,
Los Alamos, NM 87545; \\
peregrin@apo.nmsu.edu
\label{LANL1}}

\altaffiltext{2}{New Mexico State University, Department of Astronomy, 
P.O. Box 30001, 
Dept 4500, Las Cruces, NM 88003
\label{NMSU}}

\altaffiltext{3}{Department of Astronomy,
University of Washington, Box 351580,
Seattle, WA 98195
\label{UW}}

\altaffiltext{4}{Los Alamos National Laboratory, ISR-4,
MS D448, Los Alamos, NM 87545 
\label{LANL2}}

\altaffiltext{5}{Department of Physics \& Astronomy, 
University of Wyoming, 1000 E. University Blvd., 
Laramie, WY 82071 
\label{Wyo}}

\altaffiltext{6}{Apache Point Observatory, 
2001 Apache Point Road, Sunspot, NM 88349
\label{APO}}


\begin{abstract}
Multi-epoch imaging of the Orion equatorial region by the Sloan
Digital Sky Survey has revealed that significant variability
in the blue continuum persists into the late-M spectral types, indicating
that magnetospheric accretion processes occur below the substellar
boundary in the Orion OB1 association. We investigate the strength
of the accretion-related 
continuum veiling by comparing the reddening-invariant colors
of the most highly variable stars against those of main sequence
M dwarfs and evolutionary models. A gradual decrease in the
$g$ band veiling is seen for the cooler and less massive members,
as expected for a declining accretion rate with decreasing mass.
We also see evidence that the temperature of the accretion shock
decreases in the very low mass regime,
reflecting a reduction in the
energy flux carried by the accretion columns.
We find that the near-IR excess attributed to circumstellar disk thermal
emission drops rapidly for spectral types later than M4. This is likely
due to the decrease in color contrast between the disk and the cooler stellar
photosphere. Since accretion, which requires a substantial
stellar magnetic field and the presence of a circumstellar disk, is
inferred for masses down to 0.05 M$_\odot$
we surmise that brown dwarfs and low mass stars share a common mode
of formation.
\end{abstract}


\keywords{circumstellar matter -- 
stars: formation --
stars: late-type --
stars: low-mass, brown dwarfs -- 
stars: pre-main-sequence}

\section{Introduction}
Brown dwarfs are defined as objects that have less than the
minimum mass required for stable Hydrogen fusion. 
In the absence of rotation and for solar metallicity this mass,
referred to as the hydrogen-burning limit,
is 0.075 M$_\odot$ or 79 M$_{Jup}$ \citep{cha00}.  Young brown dwarfs are 
fully convective with photospheric
temperatures comparable to M stars. In the models of 
\citet{bar98} 2 Myr old proto-brown dwarfs of masses 0.020 M$_\odot$
up to the hydrogen-burning limit have temperatures
between 2500 K and 2900 K, corresponding to spectral types
of M8.5 to M6 \citep{luh03}.
The presence of a circumstellar disk in combination with the
ability of fully convective objects to generate significant magnetic fields 
should result in young brown dwarfs exhibiting Classical T Tauri behavior 
\citep{jay02}. 

The young brown dwarfs in this paper are found as isolated
objects within the Orion OB1 star formation region. Distances
and inferred ages of the Orion OB1a,b,c, and d subassociations are given
in Table \ref{tbl-oriob1} (from \citet{she03}). 
There are two possible scenarios for brown dwarf formation: (1) individual 
formation out of isolated low-mass molecular cores, just as stars are 
presumed to form from higher-massed cores, and (2) ejection of the lowest 
massed objects from a cluster though dynamical interactions.  In the 
latter scenario, objects are ejected before accreting enough material to 
grow into stars, thereby forming brown dwarfs
 \citep{rei01, kro03}.
 
Consequences of the ejection process include a broader spatial distribution,
increased velocity dispersion for the lower massed objects, and a truncation
of the circumstellar disk via tidal forces. The
ejection velocity dispersion is of order 5 km/sec corresponding to
a dispersion in proper motion of 2.6 mas/yr at
the 400 pc distance of Orion. Models of the ejection process \citep{arm97}
predict that the disk will be truncated to an outer radius of 10 AU or less,
resulting in disk lifetimes of less than 1 Myr. In order to differentiate
the star-like and ejection formation scenarios we examine 
observational signatures, such as Classical T Tauri behavior,
that are due to the presence of a circumstellar disk.

\subsection{Accretion Signatures in T Tauris}

T Tauri variables were originally identified by \citet{joy45} as a 
class of irregular 
variable stars
exhibiting marked changes in brightness and color on timescales of hours 
to days.
These changes are greatest in the blue and ultraviolet \citep{her94},
and are expected to be related to disk accretion.
Magnetospheric accretion models \citep{kon91} predict that the 
inner circumstellar disk is truncated by the magnetic field of the star.
For accretion to occur the truncation must be at or within the Keplerian
co-rotation radius, at which the orbital period of the disk matches that
of the star's rotation.

Disrupted disk material flows towards the star in accretion columns that
follow the magnetic field lines connecting the inner edge of the disk with
the high latitude regions of the star. As evidenced by the broad velocity
profiles observed in H$\alpha$ and other spectral features, much
of the permitted line emission arises in these columns. Other emission
lines, including He I and the broad component of the Ca II IR triplet,
plus the veiling blue and ultraviolet continua arise in the high temperature 
shock fronts at the base of the 
accretion columns near the stellar surface \citep{gul00}.
The lifetime of the classical T Tauri phase is thought to be 1 to 
10 Myr \citep{ken95}.

Protostellar magnetospheric accretion 
is one of a dynamic process characterized by 
intricate magnetic field topologies and unstable field-disk interactions.
Detailed MHD computations by \citet{rom02}, \citet{kuk03}, and \citet{von03}
indicate that even for the idealized
case of a dipole field aligned with the stellar rotation 
axis the accretion process is fundamentally unstable.   
An intensive photometric and spectroscopic campaign targeting the 
nearly edge-on Classical T Tauri AA Tau by \citet{bou03} shows clear evidence 
for large scale
instabilities developing in T Tauri magnetospheres as the magnetic field
lines are twisted by differential rotation between the star and the
inner disk. Observational support of magnetospheric accretion in AA Tau 
includes
time delays between the H$\alpha$, H$\beta$, He II line emission and the accretion shock 
generated emission consistent with free-fall from $\sim$8 stellar radii and the 
presence of two rotationally modulated hot spots.

\subsection{Accretion in Very Low Mass Stars and Brown Dwarfs}

There has been considerable effort over recent years to observe
and characterize the formation processes in young very low mass
stars and brown stars. This has included analyses of spectroscopic
signatures of accretion, veiling measurements, variability studies, 
and IR searches for circum(sub)stellar disk emission. 

\subsubsection{Emission Lines}

H$\alpha$ emission lines with widths in excess of 200 km/s have been
detected in young brown dwarfs
having masses as low
as $\sim$ 20 M$_{Jup}$ in many star formation regions including
IC 348, Taurus, Chamaeleon I, Upper Scorpius, and $\rho$ Oph
\citep{jay03, muz03, nat04, moh05, muz05}.
The mass accretion rate appears to scale
roughly as $M^2$ which is steeper than the linear relation inferred
by \citet{whi03} based on studies in Taurus-Auriga. 
In addition, the uncertainties in the determination of stellar parameters
(mass, radius) from evolutionary models of very low mass objects \citep{bar02}
during first few Myr contribute another source of uncertainty in these
relations. 
However, \citet{pad04} propose that the accretion rate
is controlled by accretion from the surrounding gas in the star formation
region rather than from the disk evolution. The resulting
Bondi-Hoyle accretion model yields trends consistent with the
\citet{muz03} relation although the validity of this hypothesis
needs to be checked through detailed observations and simulations.

\citet{nat04} find that the mass accretion rate scales exponentially
with the width of the H$\alpha$ line measured at the 10\% intensity
level
by comparing previous results in the substellar domain 
(\citet{muz03}, \citet{whi03}, and \citet{bar04}) with observations
in Chamaeleon I and $\rho$ Oph and study
of Classical T Tauri stars\citep{gul98}. The lowest detected accretion
rates are on the order of $10^{-11}$ M$_\odot$/yr.

\subsubsection{Optical Veiling}

Optical veiling measurements
of young brown dwarfs have been primarily obtained redward of 6000 {\AA}.
\citet{bar04a} find $r_{6200}$ = 1.0 and $r_{6750}$ = 0.25 from the
infilling of several TiO bands in the very low mass object
LS-RCrA 1, where $r_{\lambda}$ is the ratio of the excess emission
to the photospheric emission measured at a wavelength $\lambda$ ($\AA$).
Due to the presence of [S II], [O I], [O II], and [N II]
forbidden lines in the spectra they propose that this substellar
object is driving an outflow analogous to that which forms
the Herbig-Haro shocks seen near intermediate and low mass T Tauri 
stars.

Accretion shock models for very low mass stars \citep{muz00}
suggest that accretion rates as low as $10^{-9}$ M$_\odot$/yr would 
be detectable in the Johnson V band against an M6
photosphere. We expect a greater sensitivity to accretion rates 
when working in the Sloan Digital Sky Survey (SDSS) $g$
band due to increased brightness contrast between the 
shock and the photosphere. 

\subsubsection{Variability}
Optical photometric variability of low mass pre-main 
sequence stars is the result of four physical effects. These are 
rotational modulation of cool spots, rotational modulation of hot spots 
formed at the
base of magnetospheric accretion columns, instabilities
in the mass accretion rate, and flaring. 

The study
by \citet{her94} finds three distinct patterns of variability in
Weak-lined T Tauris (WTTS) and Classical T Tauris (CTTS). The Type I
variables are characterized by low amplitude multi-band fluctuations
on timescales of $\sim 0.5$ to 18 days. This behavior is primarily
seen in WTTS but can been detected in some WTTS. The second type of 
variation only occurs in the CTTS and is marked by
ireegular high amplitude brightness changes, especially in the 
near ultraviolet and the blue, on timescales
as short as a few hours. The Type III variations, like the Type II,
are irregular but occur on much longer timescales of days to weeks.
This last category are only seen in pre-main sequence stars
having early spectral types (AO to K1).

Type I variations 
are believed
due to rotational modulation of cool spots on the stellar surface.
These spots are a consequence of the magnetic activity and can persist
for over 100 rotational periods yielding periodic variations in
$VRI$. Flaring is often seen in the $B$ and especially $U$ bands.
WTTS only exhibit Type I variations.
When magnetospheric accretion is present, i.e. in the CTTS stage,
then Type II variations are seen. These are due to a combination
of rotational modulation of the accretion hot spots together with short
period fluctuations in the mass accretion rate. In most cases the
variations in the accretion mask the rotation signature.
\citet{her94} also identify a third class (Type III) variables such as 
RY Tau and
SU Aur that exhibit large non-periodic variation in the $V$ band 
but do not show significant veiling. This behavior is thought to
be caused by variable obscuration but is also limited to intermediate
and high-mass stars (spectral types A0 to K1);
we do not expect to find Type III variables in our low-mass sample.

In this sample of intermediate and low mass T Tauris \citet{her94}
find that the maximum amplitudes seen in Type I variables are 0.8 magnitudes
in $V$ and 0.5 magnitudes in $I$ for V410 Tau (SpT = K3). For the Type I
variables of spectral type M0 and later the $V$ band amplitudes are 
typically less than 0.3 magnitudes.

Variability studies of young very low mass stars and brown dwarfs
have primarily used the longer wavelength filters, most notably the
Johnson-Cousins $I$.
\citet{her02} utilized an intermediate band filter centered at 815.9 nm 
to study variability in the Orion Nebula Cluster and found that
 46\% of the stars with $\sigma < 0.1$ were periodic, where
$\sigma$ is the standard deviation measured over all
observations for a star. Only 24\%
having $\sigma > 0.1$ were periodic, the majority exhibited the
irregular variation characteristic of Type II variables.

In their study of IC 348 \citet{coh04}
found that the full range of variation for periodic
stars of spectral type M0-M4 or later is $< 0.6$ magnitudes in the
$I$ band. No CTTS exhibited periodic variability over
5 seasons of monitoring. \citet{coh04} also characterized the variability
in their sample by $\sigma$ finding that in their $I < 14.3$ subsample the WTTS
had $\sigma < 0.1$ and the CTTS were found to have $\sigma > 0.05$.
Only the Type II variables (CTTS) had $\sigma > 0.1$.

The survey of the Cha I star formation region in 
$R$, $i$, and $J$ by \citet{joe03}
found that the largest amplitudes in each band were 0.18, 0.14, and 0.13,
respectively.
The earliest spectral
type in the sample was M5, corresponding to a maximum mass of 
0.12 M$_\odot$. These
objects showed periodic variations, presumably due to spots, whose
peak-to-peak amplitudes drop in later spectral types.

$I$ band monitoring of the $\sigma$ Ori and $\epsilon$ Ori clusters
were performed by \citet{sch04a} and \citet{sch04b}. They found that periodic
variables had amplitudes less than 0.2 magnitudes.
The larger amplitude variables were generally non-periodic.
S Ori 45, a $\sim$ 20 Jupiter
mass member of the $\sigma$ Ori cluster (SpT = M8.5) was studied by
\citet{zap03} in the $IJ$ bands who measured a periodic peak to peak amplitude 
$\sim 0.2$ magnitudes.

\citet{cab04} find that nine out of 32 young brown dwarf candidates
(spectral types M5.5 to L2) exhibit $I$ band variability. The amplitudes
of mid-term (day-to-day) variations were as high as 0.36 magnitudes while
shorter term variations had standard deviations $\sim$
0.05 magnitudes. Correlations were found between high amplitude 
variability ($> 0.12$ magnitudes) and detection of either a near-IR excess 
or strong H$\alpha$ emission indicative of a possible accretion disk.

\subsubsection{Near-IR Excess}

The thermal emission from the inner circumstellar disk can be
detected in the near-IR $H$ and $K$ bands \citep{mey97}. For the
less massive and cooler young brown dwarfs the reduced contrast
between the disk emission and the substellar photosphere motivates
disk surveys in $L$ band. An $L$ band survey by \citet{liu03} finds
that for low mass stars and substellar objects in IC 348 and Taurus 
the disk fraction
is not dependent on mass or age and that the presence of disk
emission is associated with accretion indicators such as 
H$\alpha$ emission.

Comparison of disk models with mid-IR photometry shows that
the disks surrounding young brown dwarfs can be flared
like those around Classical T Tauri stars \citep{moh04}. The model
fits also suggest the presence of inner holes that are a few
substellar radii in size. 

\subsection{The Orion Equatorial Region}

The Orion OB1 a,b,c and d sub-associations 
have been the targets of several surveys. These include the optical
photometric and spectroscopic study of the $\sigma$ Ori cluster in
Orion OB1b by \citet{wal98}, the H$\alpha$ Kiso object prism
survey \citep{wir89}, and the ROSAT All-Sky Survey \citep{alc96}.
With the exception of the Orion Nebula
Cluster (OB1d) these are found along the celestial equator and are
considered in this study. The estimated ages span 1.7 Myr to 11.4 Myr
with distances between 330 and 460 parsecs corresponding
to distance moduli of 7.6 to 8.3 magnitudes. 

Recent imaging surveys of the Orion equatorial region include
\citet{wol96}, \citet{she03}, and \citet{bri03}.
\citet{wol96} performed $UBVR_cI_c$ photometric monitoring of 
X-ray selected stars
in the Orion OB1a and OB1b associations with an estimated completeness 
limit of $V$ = 18.5.
\citet{she03} conducted a deeper $BVR_cI_c$ survey 
covering 5 square degrees in Orion OB1b around $\sigma$ Ori, $\epsilon$ Ori, 
and $\delta$ Ori using the CTIO 0.9 and 1.5 meter telescopes. Their
adopted faint limit was $V$ = 20.5. The $B$ band photometry was not analyzed.

The CIDA-QUEST variability survey \citep{bri03} 
is acquiring multi-epoch
drift-scanned imaging in the Orion equatorial region using the 
1.0/1.5 meter Schmidt
telescope at The National Astronomical Observatory of Venezuela.
Exposures are acquired in the VRIH$\alpha$ bands at a plate scale of 1.02''
per pixel. The 10$\sigma$ limit is approximately $V = 19.7$ although in
selection of candidate pre-main sequence stars they specify
$V = 16 - 18.5$ \citep{bri05}, resulting in a minimum mass limit
of $\sim$ 0.12 M$_\odot$.

Studies of neighboring regions include \citet{bar04} who
performed a deep $R_cI_c$ survey of the $\lambda$ Orionis cluster
(d = 400 pc; 5 Myr) with the CFHT 12K camera. Their completeness limits
were 22.75 in both bands with the lowest mass confirmed cluster member
having 0.02 M$_\odot$ with a spectral type of M8.5.
Studies of NGC 2264 including S Mon and the Cone Nebula region \citep{reb02} 
and of the Orion Nebula Flanking fields \citep{reb00}
explicitly used ultraviolet and near-IR excess as disk
indicators. Their limiting magnitudes are $U \sim 20$, $V \sim 20$,
and $I_C \sim 18$.

\subsection{Program}

The goal of this paper is to study the trends in veiling emission,
variability, and near-infrared excess in the low mass stars and brown
dwarfs found in the Orion OB1a,b, and c subassociations. 
If brown dwarfs form in the same manner as stars we expect that these 
characteristics will vary smoothly across the stellar/substellar 
mass boundary.

The remainder of this paper is organized as follows:
the SDSS and Two Micron All-Sky Survey (2MASS) observations used in this paper 
are presented
in \S 2. In \S 3 we discuss the expected colors of young mass stars and 
brown dwarfs using the colors of 
spectroscopically verified field M dwarfs \citep{wes04}, 
the evolutionary models
of \citet{bar98} (hereafter BCAH98), and the bolometric corrections of 
\citet{gir04}. 
The selection criteria for young low mass objects including an
estimate on the contamination fraction are presented in \S 4.
We discuss the empirical trends in variability and color
in \S 5 and in \S 6 we provide a summary of results and upcoming work.

\section{Observations}

\subsection{Photometry}

The Sloan Digital Sky Survey (SDSS) obtains deep 
photometry with magnitude limits (defined by 95\% detection repeatability 
for point sources) of 
$u=22.0$, $g=22.2$, $r=22.2$, $i=21.2$ and $z=20.5$. These five passbands,
$ugriz$, 
have effective wavelengths of 3540, 4760, 6290, 7690, and
9250 {\AA}, respectively.
A technical summary of the SDSS is given by \citet{yor00}.
The SDSS imaging camera is described by \citet{gun98}. \citet{ive04}
discuss the data management and photometric quality assessment system.

The Early Data Release and the Data Release One are described by
\citet{sto02} and \citet{aba03}. The former includes an extensive 
discussion of the data outputs and software.  \citet{pie03} describe the 
astrometric calibration of the survey and the 
network of primary photometric standard stars is described by 
\citet{smi02}. The photometric system itself is defined by \citet{fuk96},
and the system which monitors the site photometricity by
\citet{hog01}.
\citet{aba03} discuss the differences between the native SDSS 2.5m
$ugriz$ system and the $u'g'r'i'z'$ standard star system defined
on the USNO 1.0 m \citep{smi02}.

The SDSS low Galactic latitude data which includes the Orion equatorial 
imaging used in this work are described by \citet{fin04}. 
2MASS \citep{skr97} obtained nearly
complete coverage of the sky in $JHK_s$.
The multi-epoch
data (Table \ref{tbl-runs}) we use in this study were obtained 
under photometric conditions
and cover $80\degr < \alpha_{2000} < 90\degr$ and 
$-1.25\degr < \delta_{2000} < 1.25\degr$.

\subsection{Reddening-Invariant Indices}

In this work we employ reddening-invariant indices 
of the form
$Q_{xyz} = (x-y) - (y-z) \times E(x-y)/E(y-z)$.
$Q_{xyz}$ is dependent upon the assumed ratio of general to selective
extinction ($R_V = A_V/E(B-V)$; \citet{car89}). 
Here $xyz$ refer to the specific 
passbands, e.g.
$ugrizJHK_S$ and $E(x-y)$ is the color excess due to reddening in the
$x-y$ color. This definition of reddening-invariant colors
follows the original \citet{joh53} $Q$ that would be written as 
$Q_{UBV}$ in our notation. In the $(x-y,y-z)$ color-color diagram the 
$Q_{xyz}$ axis is perpendicular to the reddening vector. 

This approach 
has previously been used to study pre-main sequence populations by
\citet{deg89} and \citet{bro94} for the Scorpius-Centaurus OB and Orion
OB1b associations, respectively.
In both cases reddening-invariant colors
were formed by $[x-y] = (x-y) - (V-B) \times E(x-y)/E(V-B)$.
using the $VBLUW$ Walraven system. Stellar parameters were 
estimated using reddening-invariant two-color diagrams to compare
observed stellar colors against those from Kurucz model grid \citep{kur79}.
  
While the contrast between the accretion
shock and the cool photosphere is greatest in the SDSS $u$ band,
we choose not to use the $Q_{ugr}$ color due to sensitivity issues and 
the presence of a ``red leak'' in the SDSS imager $u$ filter.
M dwarfs are 2.5 magnitudes fainter in $u$ than in $g$, thus use of the
$u$ band in Orion would not reach young brown dwarfs.

We use the extinction tables derived by 
D. Finkbeiner\footnote{private communication; 
see http://www.astro.princeton.edu/$\sim$dfink/sdssfilters/} 
to define the coefficients used in defining reddening-invariant
colors. These tables contain the $A_X/E(B-V)$ values for the
SDSS $ugriz$ filters for specific values of $R_V$ and source
spectra. The values we present here are obtained using
an F dwarf source spectrum and $R_V$ = 3.1 and 5.5. The $R_V = 3.1$
case is the standard extinction law found in the diffuse ISM. 
The $R_V = 5.5$
law is representative of that found in molecular clouds due to 
larger dust grains and is shown here for example.
Throughout this paper we will adopt 
$R_V = 3.1$ as it is characteristic of all but perhaps the most
heavily extincted regions of the Orion OB1b association
\citep{war78}.

\begin{equation}
Q_{gri} = \cases{
(g-r) - 1.852 (r-i) & \rm{$R_V$ = 3.1}, \cr
(g-r) - 1.339 (r-i) & \rm{$R_V$ = 5.5}  \cr
}
\end{equation}
and
\begin{equation}
Q_{riz} = \cases{
(r-i) - 0.987 (i-z) & \rm{$R_V$ = 3.1}, \cr
(r-i) - 1.004 (i-z) & \rm{$R_V$ = 5.5}. \cr
}
\end{equation}
The near-IR reddening-invariant color is independent of the
value of $R_V$, from \citet{sch98} we obtain
\begin{equation}
%\begin{split}
Q_{JHK} = (J-H) - 1.563 (H-K).
%\end{split}
\end{equation}

\section{Expected Colors of Young Brown Dwarfs}

\begin{figure}
\plotone{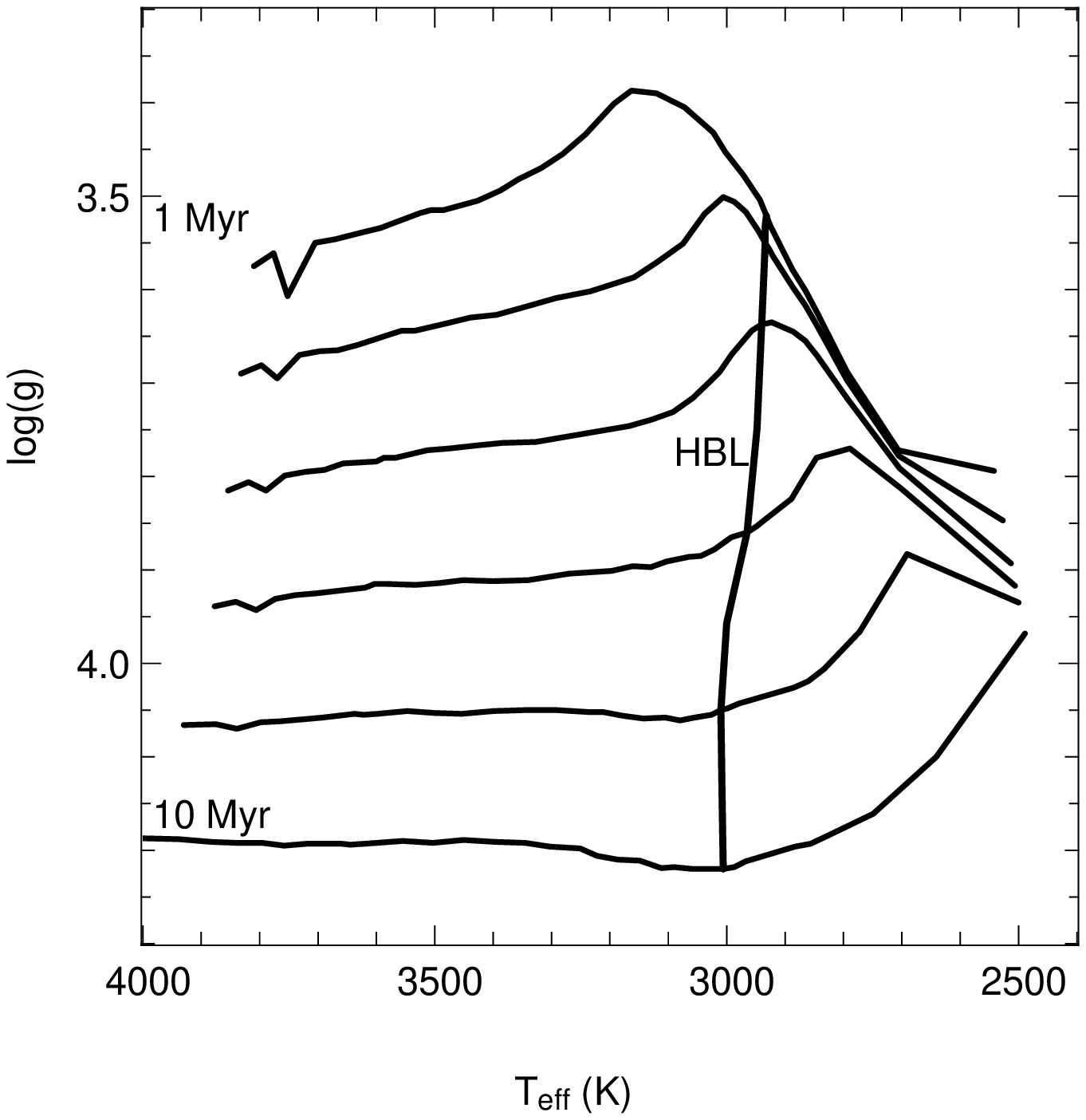}
\caption{{\bf Temperature and surface gravity for young low mass stars.} 
The predicted effective temperatures (K) and 
surface gravities (log(g)) are shown here for the BCAH98 models having 
$0.02 M_\odot \le M \le 1.0 M_\odot$ 
and ages between 1 Myr and 10 Myr. 
Isochrones are
drawn for log(years) = 6.0 to 7.0, spaced by 0.2 dex. The thick black
line traces the Hydrogen Burning Limit ($0.075 M_\odot$). 
\label{fig-teff_logg}}
\end{figure}

In this section we compare the expected colors of young
brown dwarfs against those of field M dwarfs of equivalent
temperatures. Examination of the temperature - surface gravity (Figure
\ref{fig-teff_logg}) 
relations based on the BCAH98 models
shows for objects at the Hydrogen Burning Limit younger than 10 Myr the 
effective temperature ranges between 2900 K and 3000 K
and the surface gravity varies from log(g) = 3.5 to 4.2. 

When the locations in Hertzsprung-Russell diagrams of spectroscopically 
classified
and presumably co-eval stars in specific star formation regions are compared
against evolutionary models the inferred ages mismatch using the 
spectral type to $T_{eff}$ scale defined for disk dwarfs. This has led to
the adoption of a semi-empirical temperature scale intermediate between 
that of the dwarfs and the giants \citep{luh03}. This practice is motivated
by the surface gravities expected
in pre-main sequence stars and results in effective temperatures
several 100 K warmer at a given spectral type. Thus a $T_{eff}$ of 2900 K 
corresponds to a spectral type of M4.5 for a dwarf \citep{rei00} but of 
M6.5 for a pre-main sequence star.

The photospheric colors of M dwarfs and M giants differ at
the later spectral types with the M giants becoming bluer due
to surface gravity effects. 
This color shift with lower surface
gravity is due to the change in the optical depth
for the formation of molecules, most importantly TiO, that dominate the 
stellar spectrum. \citet{luh99} notes that for objects of later spectral
types, e.g. M8, the trend is reversed with the lower surface gravity stars 
becoming redder longward of 8500 {\AA}.
Due to these deviations from disk dwarf colors 
UV-excess and blue-excess techniques based on the ZAMS become 
ambiguous for pre-main sequence stars later than M4 \citep{reb00}.

\subsection{Disk Dwarf Colors}

\begin{figure}
\plotone{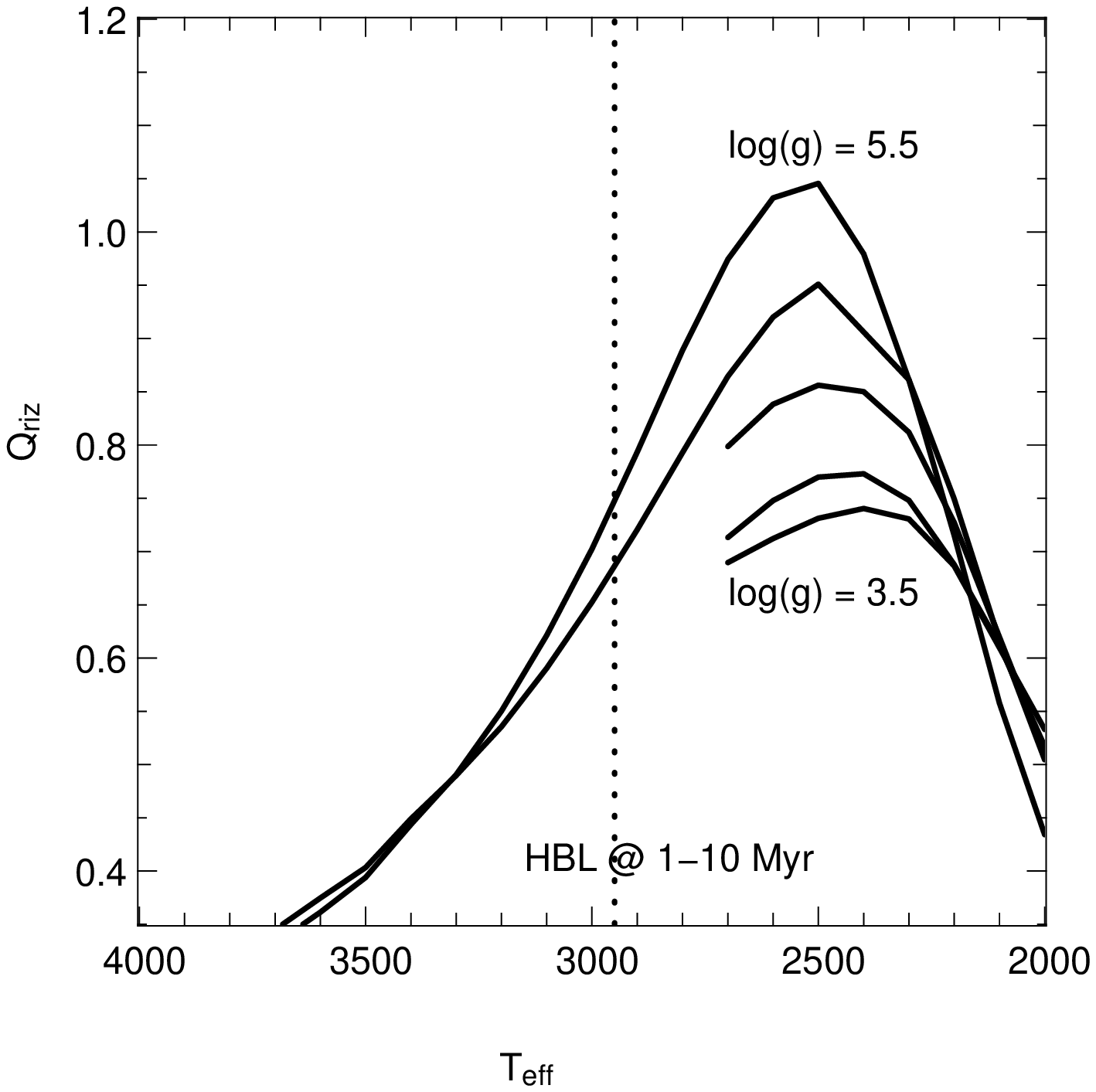}
\caption{{\bf Model $Q_{riz}$ color and $T_{eff}$ relations}. 
This figure shows the predicted $Q_{riz}$ colors as a function
of effective temperatures using the $ugriz$ bolometric 
correction tables of \citet{gir04} which are based on the
\citet{all00} model atmospheres. The lower surface gravity
models (log(g) = 3.5, 4.0 and 4.5) were not presented by 
\citet{gir04} for $T_{eff} > 2700$. The vertical line marks the 
approximate location of the Hydrogen Burning Limit for the
first 10 Myr.
\label{fig-teffqriz}}
\end{figure}

Figure \ref{fig-teffqriz}, based on the sample of \citet{wes04},
shows that the $Q_{riz}$ color index increases towards
cooler spectral types among M dwarfs,
reaching a maximum
value $\sim 1.2$ at M8. At the ages in which we expect to see
accretion activity this turnover occurs for objects near
0.02 $M_\odot$ which are too faint to be included in this survey. 

To study the
change in $Q_{gri}$ relative to the disk dwarf locus we obtain 
linear fits between $Q_{gri}$ and $Q_{riz}$ using photometry of M dwarfs in
the SDSS Third Data Release (DR3; \citet{aba05}) obtaining
\begin{equation}
Q_{gri}  = \cases{
-3.371 Q_{riz} + 1.119 & \rm{$R_V$ = 3.1}, \cr
 -2.241 Q_{riz} + 1.163 & \rm{$R_V$ = 5.5} \cr
}
\end{equation}
using a minimum threshold to $Q_{riz}$ of 0.5 to
avoid the slope change in the color-color diagram at the 
earliest M dwarfs. The fit residuals have RMS values of
0.148 and 0.117 for $R_V$ = 3.1 and 5.5, respectively.

\begin{figure}
\plotone{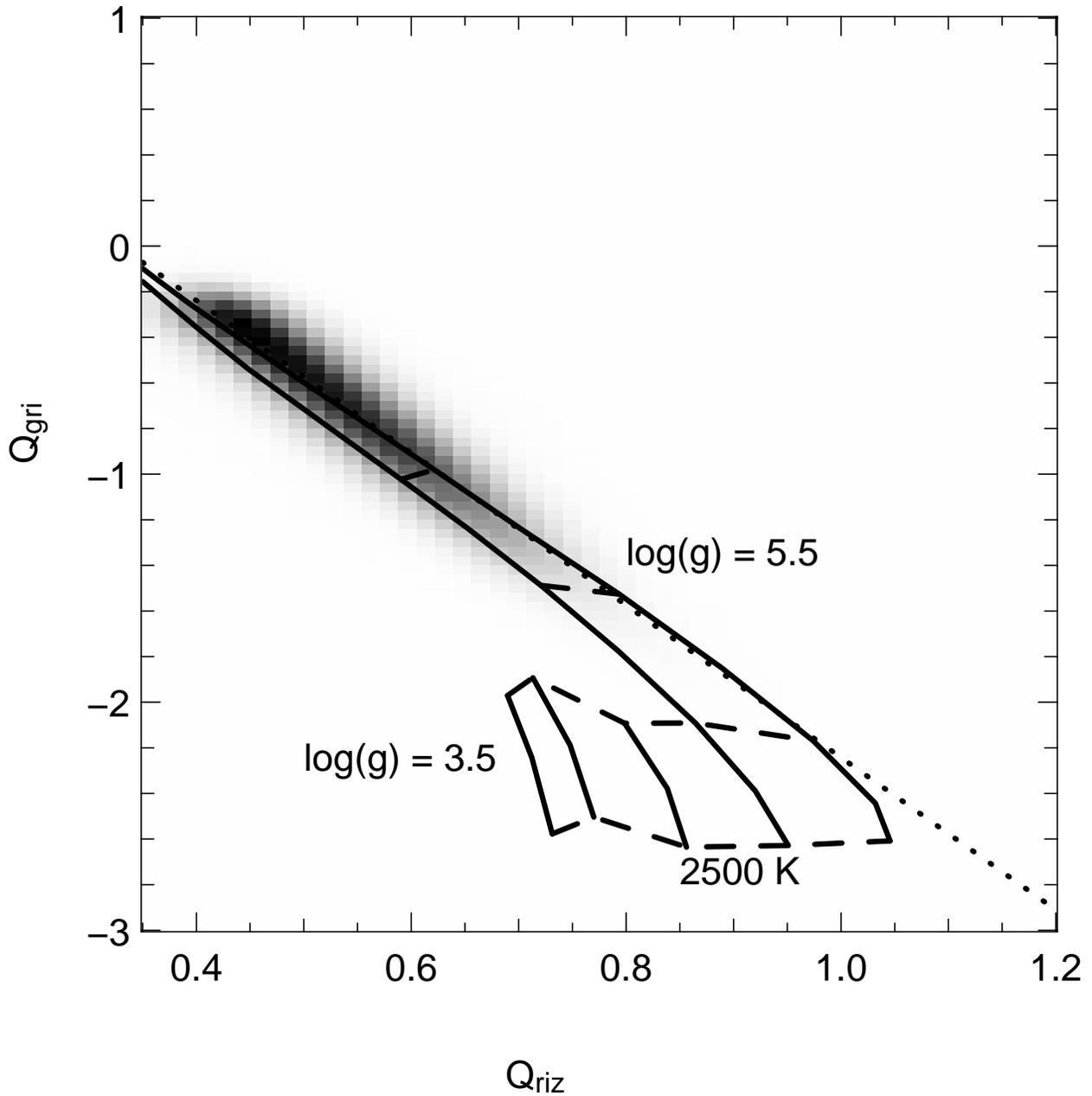}
\caption{{\bf Disk dwarf reddening-invariant color-color relations.}
This figure shows ($Q_{riz}$, $Q_{gri}$) color-color diagrams
for 1,116,184 M star candidates in the SDSS Third Data Release
\citep{aba05}. The dotted
diagonal line traces linear fits based on the $Q_{riz} > 0.5$ subsample
of 694,062 stars. The \citet{gir04} grid is overlaid with
dashed lines connecting models having $T_{eff}$ = 2500 K, 2700 K, 2900 K,
and 3100 K. 
\label{fig-invarcc}}
\end{figure}

The $R_V$ reddening-invariant two-color diagram is shown
in Figure \ref{fig-invarcc} with the theoretical isochrones
of \citet{gir04} overplotted.  These isochrones are based on the 
``AMES'' 
(4000 K $\ge T_{eff} \ge$ 2800 K) and ``AMES-dusty'' 
(2800 K $> T_{eff} \ge$ 500 K)
model atmospheres
of \citet{all00} for cool stars and brown dwarfs. While the latter
cover log(g) between 3.5 and 6.0, the $ugriz$ bolometric corrections were
only computed for the warmer ``AMES'' models having log(g) of 5.5 and 6.0.

\begin{figure}
\plotone{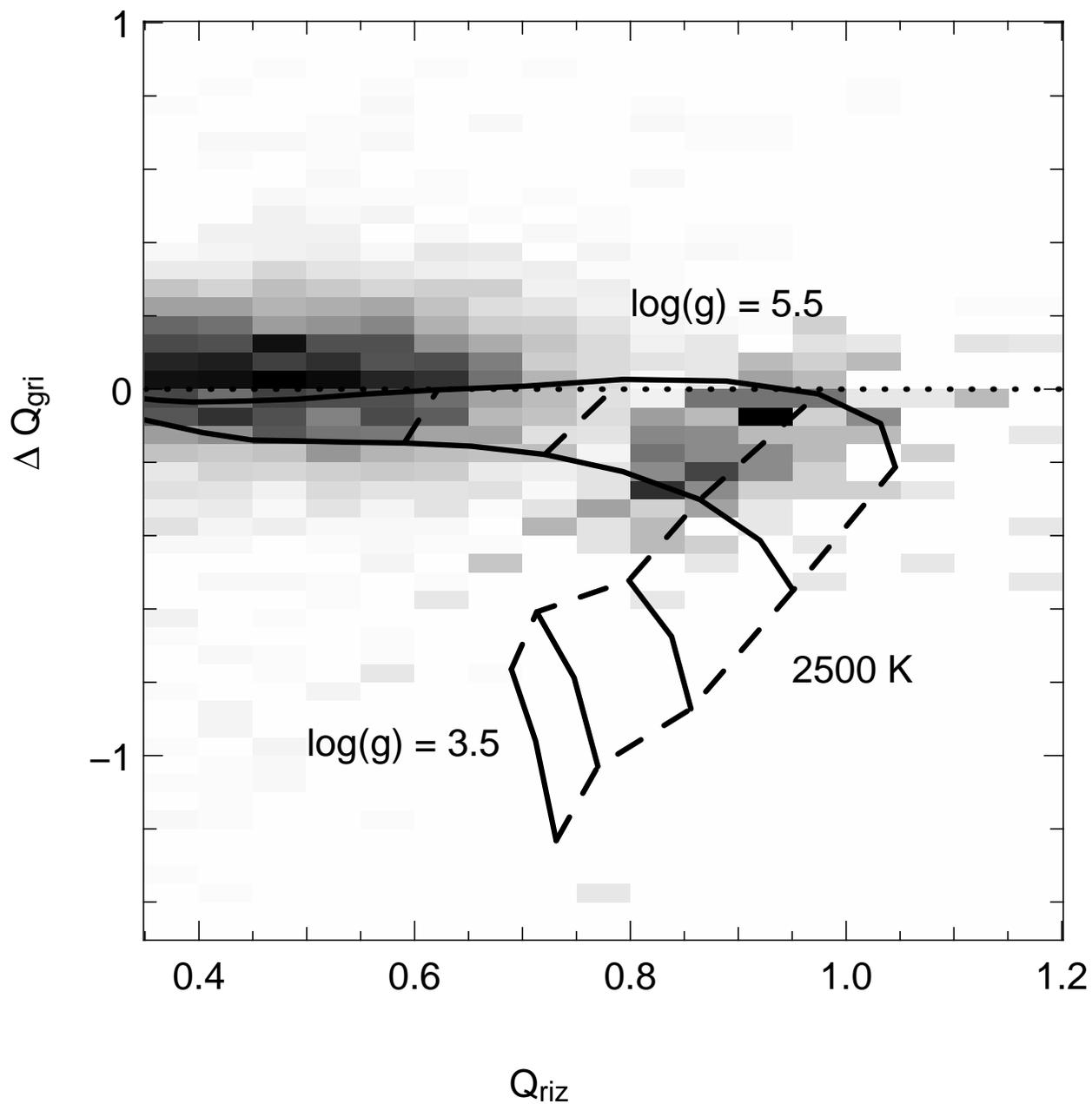}
\caption{{\bf Fit residuals for SDSS M dwarfs.}
The fit residuals ($\Delta Q_{gri}$) are plotted against $Q_{riz}$
for the \citet{gir04} model isochrones and the SDSS spectroscopic sample 
of \citet{wes04}. The meaning of the lines follows Figure
\ref{fig-invarcc}.
For the purposes of illustration this M dwarf sample is subdivided
at M4.5 with the resulting 2-D histograms scaled to match peak counts.
\label{fig-daqriz}}
\end{figure}

The fit residuals ($\Delta Q_{gri}$) are plotted against $Q_{riz}$
for the model isochrones and the spectroscopic sample of \citet{wes04}.
For the purposes of illustration this M dwarf sample is subdivided
at M4.5 with the two resulting
histograms in color-color space scaled to match peak counts. As is
seen in Figure \ref{fig-daqriz} the observed reddening-invariant
colors are well-described by the log(g) = 5.0 and 5.5 models appropriate
for the disk dwarfs. The blueward trends in both $Q_{riz}$ and
$\Delta Q_{gri}$ increasing with decreasing surface gravity as shown
by the log(g) = 4.5, 4.0, and 3.5 models. The dotted lines connect models
of $T_{eff} = $ 2500 K, 2700 K, 2900 K, and 3100 K.
We refer to $\Delta Q_{gri}$ as the observed $Q_{gri}$ offset since
it is measured relative to the $(Q_{riz},Q_{gri})$ linear fit. As described
below, understanding
the effect of low surface gravity on $\Delta Q_{gri}$ is crucial for
estimating the $g$ band veiling from reddening-invariant colors.

\subsection{Veiling Models}

The identification of accretion shocks by photometry involves
detection of the additional continuum flux generated by the
reprocessing of the high temperature shock back along the 
accretion column. The veiling produced by this excess continuum
is defined as the ratio of the shock flux to that of the stellar 
photosphere within a specified wavelength range or filter passband.

The observed
increase in brightness in an arbitrary filter $x$ is
\begin{equation}
\label{eqn-veiling}
%\begin{split} 
\Delta m_x  = -2.5 {\rm log}(1 + r_x)
%\end{split}
\end{equation}
where $r_x$ is the veiling in $x$ and $m_x$ is the observed
magnitude.
The observed veilings in low mass T Tauris are significantly higher
in the near-ultraviolet and blue regions of the spectrum than 
in the red. For all but the mostly heavily veiled sources the
contribution of the accretion continuum to the photosphere is
minimal at wavelengths above 7000{\AA} \citep{muz03} and therefore
should be minimal in the SDSS $i$ and $z$ bands. This is a contrast
effect between the $\sim 10^4$ K shock emission and the $\lesssim$ 3000 K
star. 

The change in a reddening-invariant color due to veiling is then
\begin{eqnarray}
%\begin{split}
\Delta Q_{xyz} &  = & -2.5 \{{\rm log}[(1 + r_x)/(1 + r_y)] + \nonumber \\
&& {\rm log}[(1 + r_y)/(1 + r_z)] E(x-y)/E(y-z)\}.
%\end{split}
\end{eqnarray}
Our veiling models are characterized by the veilings in the
SDSS $g$ and $r$ band,
$r_g$ and $r_r$. We explicitly assume that $r_i$ and $r_z$ are zero
({\it cf. supra}) and obtain 
\begin{equation}
%\begin{split}
\Delta Q_{gri}  =  -2.5 \{{\rm log}[(1 + r_g)/(1 + r_r)] -  {\rm log}[1 + r_r] E(g-r)/E(r-i)\}
%\end{split}
\end{equation}
and 
\begin{equation}
%\begin{split}
\Delta Q_{riz} = -2.5 \{{\rm log}[1 + r_r]\}.
%\end{split}
\end{equation}

\begin{figure}
\plotone{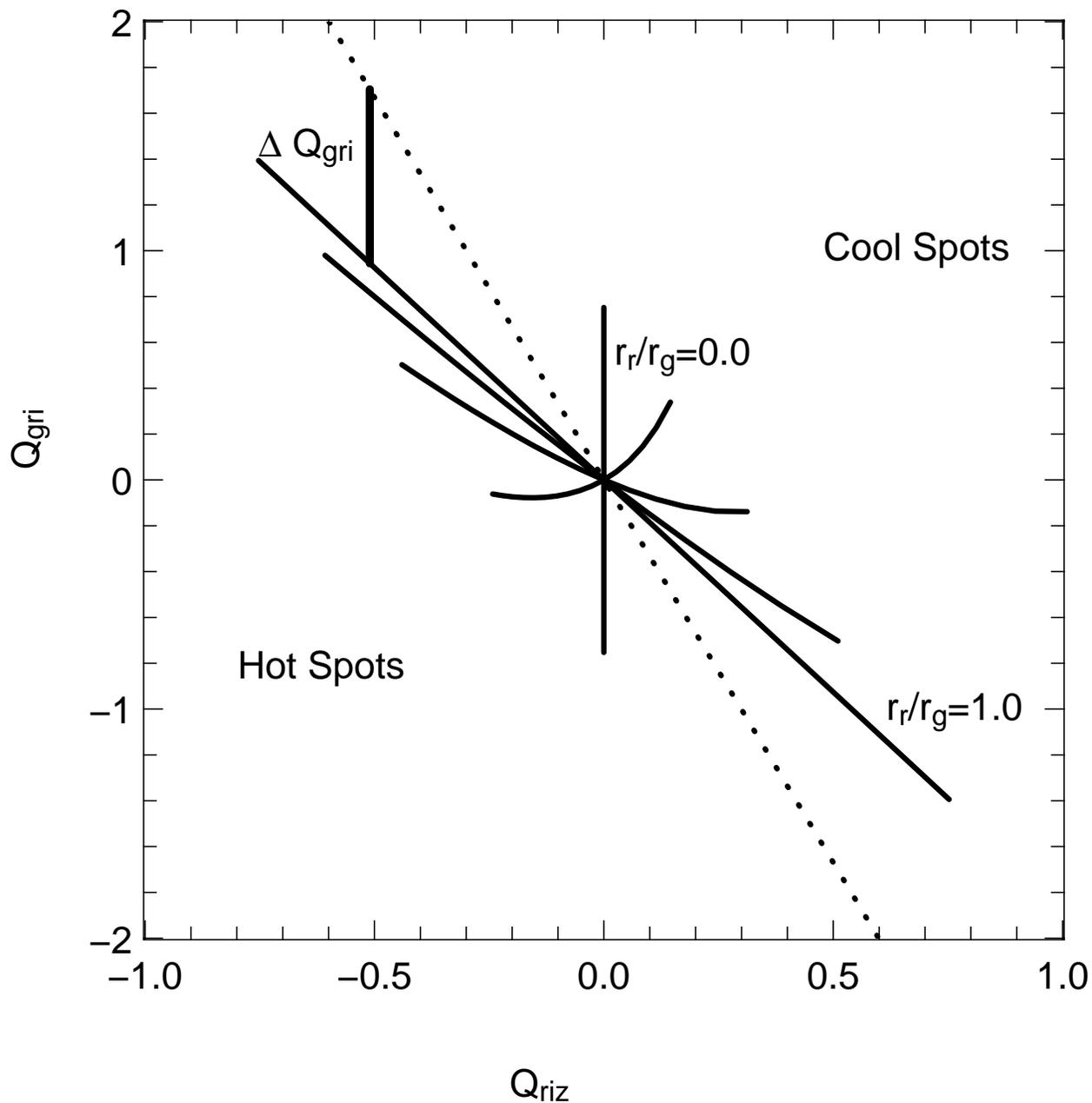}
\caption{{\bf Color changes due to veiling.} 
The changes in $Q_{riz}$ and $Q_{gri}$ are shown
for the ratio of $r_r$ to $r_g$ varying from 0.0 to 1.0
in steps of 0.25 assuming a $R_V$ = 3.1 reddening law.
The $g$ band veiling varies from -0.5 to 1.0 along each curve.
The dotted line shows the linear fit to the disk dwarf locus
assumed in computation of $\Delta Q_{gri}$.
\label{fig-model_dqxyz}}
\end{figure}

In the general case where $r_r > 0$ we can not deduce the
photospheric spectral type from $Q_{riz}$.
However, since veiling makes stars look bluer (earlier, more
massive), we can use the observed $Q_{riz}$ to deduce the earliest 
possible spectral type.
For veiling models consisting of a non-zero $r_g$ and $r_r$ 
we can examine the variations in the $Q_{riz}$ and $Q_{gri}$ color.
Figure \ref{fig-model_dqxyz} shows the
color shifts for $r_g$ varying continuously from -0.5 to 1
and $r_r/r_g$ = 0.00, 0.25, 0.50, 0.75, and 1.00. The negative veiling
values are used to model the cool spots seen in Weak-lined
T Tauris.

\begin{figure}
\plotone{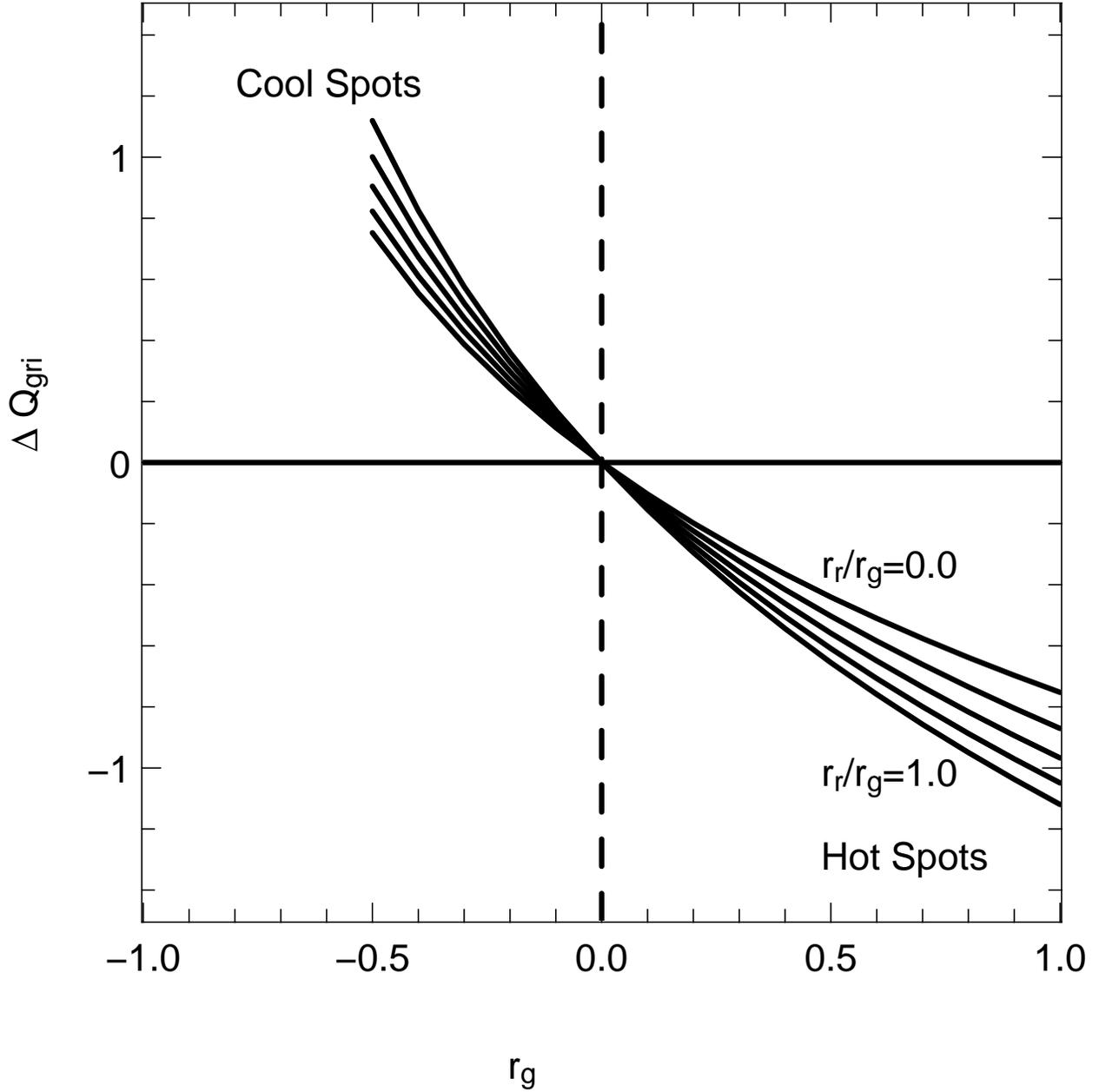}
\caption{{\bf Relation between observed $\Delta Q_{gri}$ and veiling.} 
The observed vertical separation ($\Delta Q_{gri}$) from the dwarf locus
for stars veiled in both the $g$ and $r$ bands is shown here against the
$g$ band veiling ($r_g$). The models used are those shown in Figure \ref{fig-model_dqxyz}.
\label{fig-model_dobsqxyz}}
\end{figure}

When an unknown veiling is present in the $r$ band, the key
diagnostic feature in the $(Q_{riz}, Q_{gri})$ color-color diagram is the
vertical displacement of the star from the stellar locus along the
$Q_{gri}$ reddening-invariant index.  This is
shown in Figure \ref{fig-model_dobsqxyz} as a
function of $r_g$ assuming the linear fits to the locus given
above. In the case of weak to moderate veiling, $r_g < 1$, 
we see that differing the $r_r$ to $r_g$ ratio only results
in a minimal spread of derived $r_g$ for a given $\Delta Q_{gri}$. 
We can therefore use
the observed $\Delta Q_{gri}$ as a proxy for continuum veiling
even when the spectral type and $Q_{riz}$ color are not
known {\it a priori} assuming that the intrinsic $Q_{gri}$ offset
due to low surface gravity can be inferred.

\section{Candidate Selection}

Possible low mass and very low mass accreting stars were selected
on the basis of color, location in color-magnitude diagrams, and
variability. An estimate of contamination was obtained by applying
these criteria to a neighboring field.

\subsection{Color and Magnitude Criteria}
 
We adopted selection criteria $(r-i) > 0.6$ and $Q_{riz}> 0.35$
to target M dwarf colored stars. This corresponds to $T_{eff} <$ 3700 K
or spectral types later than M0.5 in the field and M1 on the \citet{luh03} 
scale. The maximum stellar mass inferred from
the 2 Myr \citet{bar98} isochrone is 0.8 M$_\odot$.
 
To reduce the scatter in the reddening 
invariant colors we required photometric errors less than 0.05 magnitudes 
in $grizJH$.
This was relaxed for the $K_S$ band to extend the detection limit.
The resulting magnitude and error limits are summarized
in Table \ref{tbl-limits}. The limiting $J$ band magnitude of 15.5
corresponds to a lower mass limit of 0.05 M$_\odot$ ($\sim$ M7)
based on the BCAH98 2 Myr isochrone.
Propagation of these errors into the
expressions for the reddening-invariant colors yields maximum
errors for $Q_{gri}$ of 0.088 and 0.075 and for $Q_{riz}$ of
0.045 and 0.046, each in the cases of $R_V$ = 3.1 and 5.5,
respectively. The maximum error in $Q_{JHK}$ is 0.159 which is
dominated by the $K_S$ error of 0.1 at the faint limit.

In order to remove background stars we required that candidates
were brighter than the main sequence placed at the distance
of Orion in both the $(r-i,r)$ amd $(i-z,i)$ color-magnitude diagrams.
The color-absolute magnitude relations in the SDSS passbands for 
late K and M dwarfs are $M_r = (6.110 \pm 0.418) + 
(3.800 \pm 0.320) (r-i)$ and $M_i = (5.936 \pm 0.496) +
(6.308 \pm 0.803) (i-z)$ (Golimowski et al. 2005 {\it in prep.}).
The errors in these relations are dominated not by photometric
uncertainly (which is typically 1-2\%), but by the intrinsic cosmic scatter
of the sample.

\subsection{Variability}

Due to the non-periodic variations of
low mass CTTS and the irregular sampling by the SDSS we expect the
resulting lightcurves to be fairly random. A truly random sequence
has a standard deviation ($\sigma$) equal to 1.0/$\sqrt{12}$ of the 
peak-to-peak amplitude. 
Given a finite number of samples and a probability that the true value of
$\sigma$ exceeds a specific
threshold we need to find the corresponding minimum threshold for
the observed $\sigma$  that indicates significant variability.
We generated a series of Monte Carlo simulations to determine the maximum
$\sigma$  in a random sequence given an intrinsic $\sigma$ 
of the parent population
and a limited number of observations. The scaling factors presented in
Table \ref{tbl-varthres} 
reflect the maximum $\sigma$ seen in 99\% of the simulated data sets using 
10,000 realizations per point. In the limit of large number of observations 
the scaling factors tend towards unity. 

We identify variable stars on the basis of the $g$ band $\sigma$ exceeding
both a threshold of 0.05 magnitudes (scaled by the correction factor defined 
above), and
three times the photometric error in $g$. Of these we also select a subset
that meet the same conditions applied to the $z$ band. This 
$z$ band variability criterion matches the typical boundary
between Type I and Type II $I$ band variations in very low mass stars and 
young brown dwarfs.

\subsection{Results}

\begin{figure}
\plotone{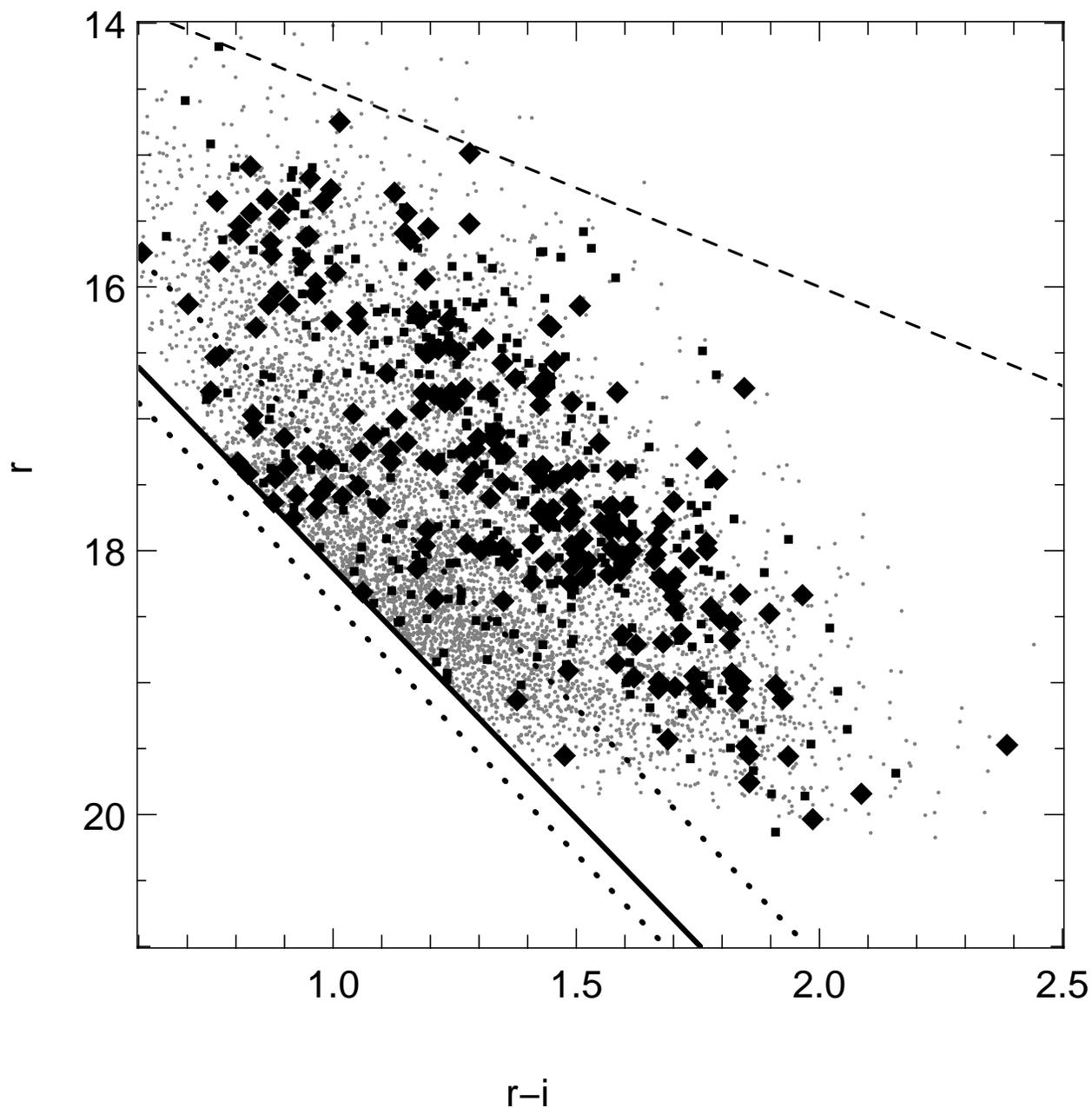}
\caption{{\bf $(r-i,r)$ color-magnitude diagram.}.
The Orion PMS locus is traced by the 
variable stars ({\it squares}: $g$ band $\sigma > 0.05$ magnitudes,
{\it diamonds}: both $g$ band $\sigma$ and $z$ band $\sigma > 0.05$ 
magnitudes.) 
in this
color-magnitude diagram. The solid line indicates
the main sequence at the $m-M = 8.22$ distance of Orion OB1b.
The approximate survey bright limit based on saturation in the
$z$ band is marked by the dashed line and the distance to the 
background Gould Belt population is delimited by the two dotted lines.
\label{fig-ori_rricmd}}
\end{figure}

\begin{figure}
\plotone{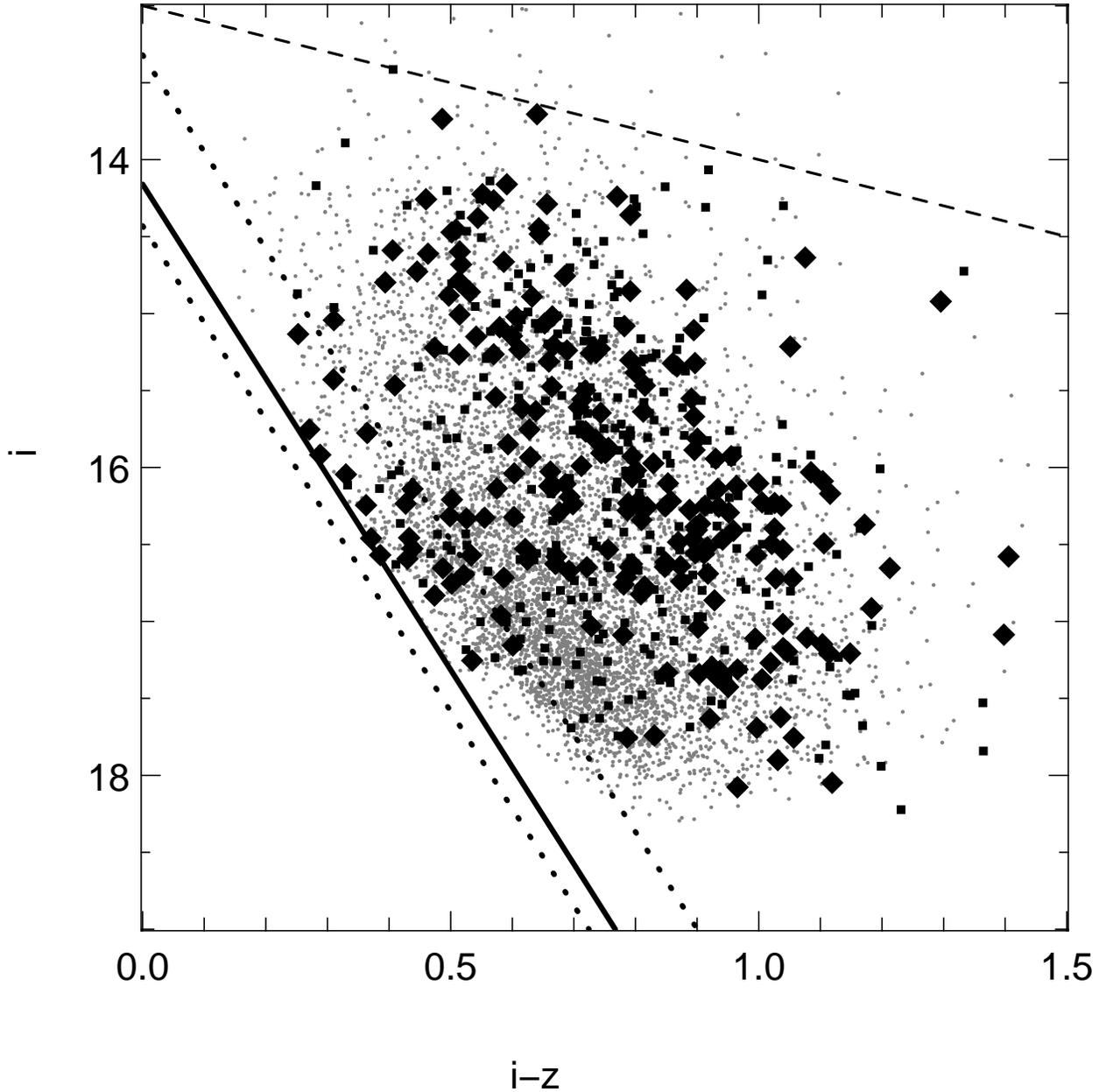}
\caption{{\bf $(i-z,i)$ color-magnitude diagram.}.
The Orion PMS locus is traced in this color-magnitude diagram
following Figure \ref{fig-ori_rricmd} above. 
The solid line indicates
the main sequence at the $m-M = 8.22$ distance of Orion OB1b.
The approximate survey bright limit based on saturation in the
$z$ band is marked by the dashed line and the distance to the 
background Gould Belt population is delimited by the two dotted lines.
\label{fig-ori_iizcmd}}
\end{figure}

\begin{figure}
\plotone{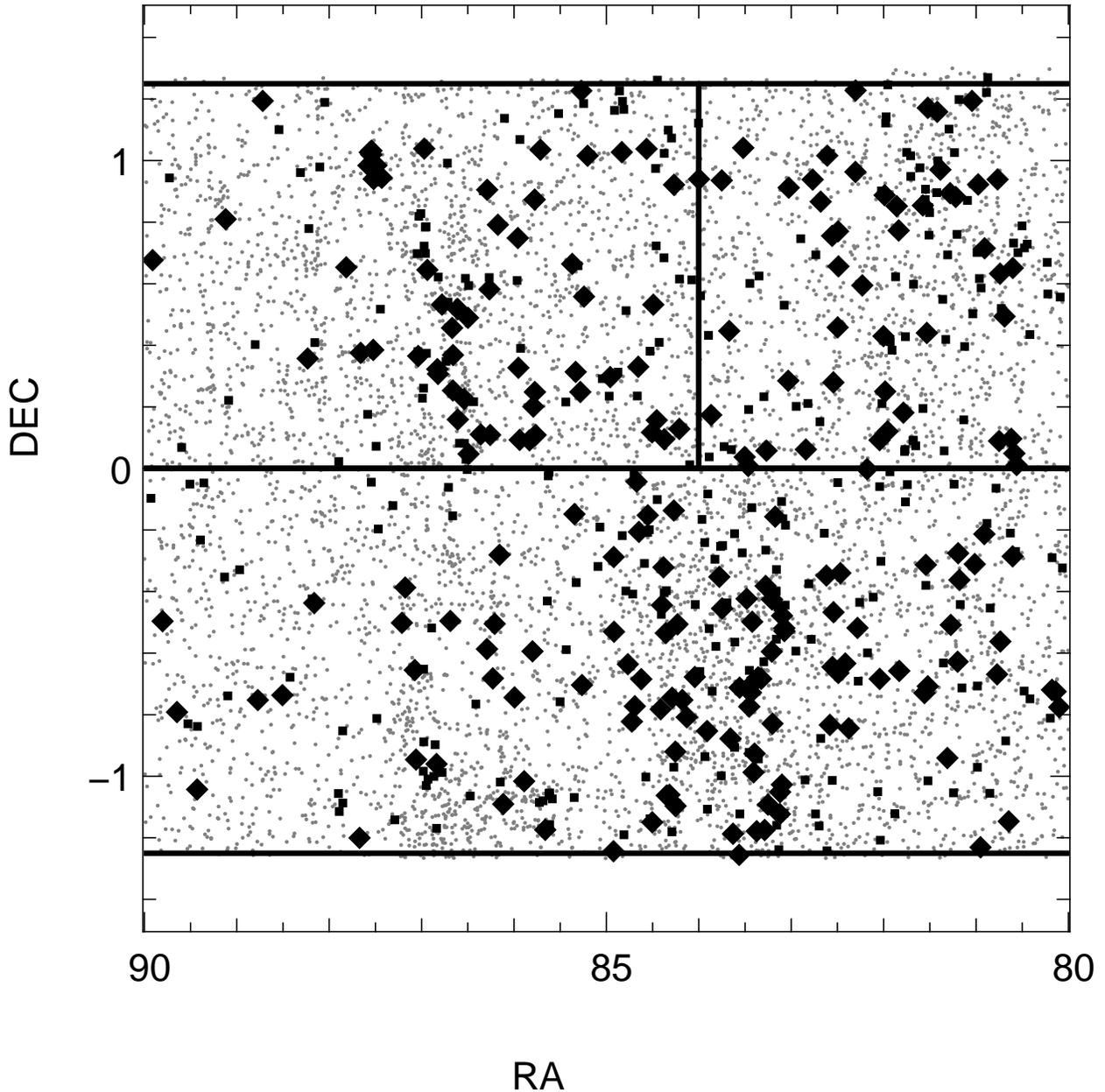}
\caption{{\bf Orion spatial distribution.} 
The distribution in RA and DEC is shown here for the
pre-main sequence candidates in the
Orion sample. The variable stars are marked following Figure 
\ref{fig-ori_rricmd}.
A portion of the Orion OB1b assocation is
visible for $82\degr < \alpha_{2000} < 85\degr$. The NGC 2068/2071
protocluster is seen at $\alpha_{2000} = 86.5\degr$. The cluster
at $\alpha_{2000} \sim 81\degr$, $\delta_{2000} \sim 1\degr$ is part of the
Orion OB1a association. The lines trace the approximate boundaries
of the portions of the 
L1630 cloud ({\it upper left}), the Orion OB1a association
({\it upper right}), and the Orion OB1b association ({\it lower half})
included in this study.
\label{fig-ori_radec}}
\end{figure}

On the basis of these criteria we identify 507 stars that are 
significantly variable in the $g$ band of which 215 also meet the criteria
for $z$ band variability. 
In Figures \ref{fig-ori_rricmd} and \ref{fig-ori_iizcmd}
we see that the majority of the variable
stars form a locus parallel to and roughly
2 magnitudes above the main sequence placed at the $m-M = 8.2$ distance
of the Orion star formation region. 
The spatial distribution of these low mass candidates is 
concentrated in the Orion OB1a and OB1b 
associations 
and the NGC 2068/2071 protocluster (Figure \ref{fig-ori_radec}).
The Orion OB1c subassociation members comprise a dispersed
population in this region. 

\subsection{Comparison Field}

In order to assess the possible contamination of our Orion sample by variable
field stars we apply the criteria defined above to a neighboring equatorial
field bounded by $40\degr < \alpha < 70\degr$ and $-1.25\degr < \delta < 1.25\degr$ which
encompasses three times the area of our survey. This is a complex region
containing at least three populations of young stars which we now describe.

On the basis of ROSAT All-Sky Survey \citep{tru83} detections \citet{neu97}
subsequently identified a population of $\sim$ 30 Myr low mass WTTS 
south of the Taurus-Auriga cloud. While we expect to see Class I variability
in these stars signs of active accretion or the presence of inner circumstellar
disks should not be evident due to the typical 10 Myr disk lifetime. 
We also expect to detect the background 8 Myr old Gould Belt stars 
\citep{gui98} which may include CTTS. As the Gould Belt midplane crosses
our equatorial survey area near $\alpha = 70$ we suspect some Gould Belt 
stars might be found in our Orion sample.

The third star formation complex is that associated with the nearby
(80 $\pm$ 20 pc) translucent cloud MBM 18 \citep{lar03}. This
high latitude cloud contains compact regions of CO(1-0) emission with 
line intensities above 3 K. If MBM 18 is an active star formation
region then due to its proximity we expect all but the lowest massed
young stars to be too bright for the SDSS,

Application of our Orion CTTS criteria results in 133 $g$ band variables of
which 19 are also variable in the $z$ band. After scaling by the ratio
of the survey areas these imply possible contamination fractions of
8.7\% and 2.9\% for the Orion $g$ band variables and its $z$ band variable 
subset, respectively.

\section{Empirical Results For The Orion Population}

\subsection{Veiling Signatures}

\begin{figure}
\plotone{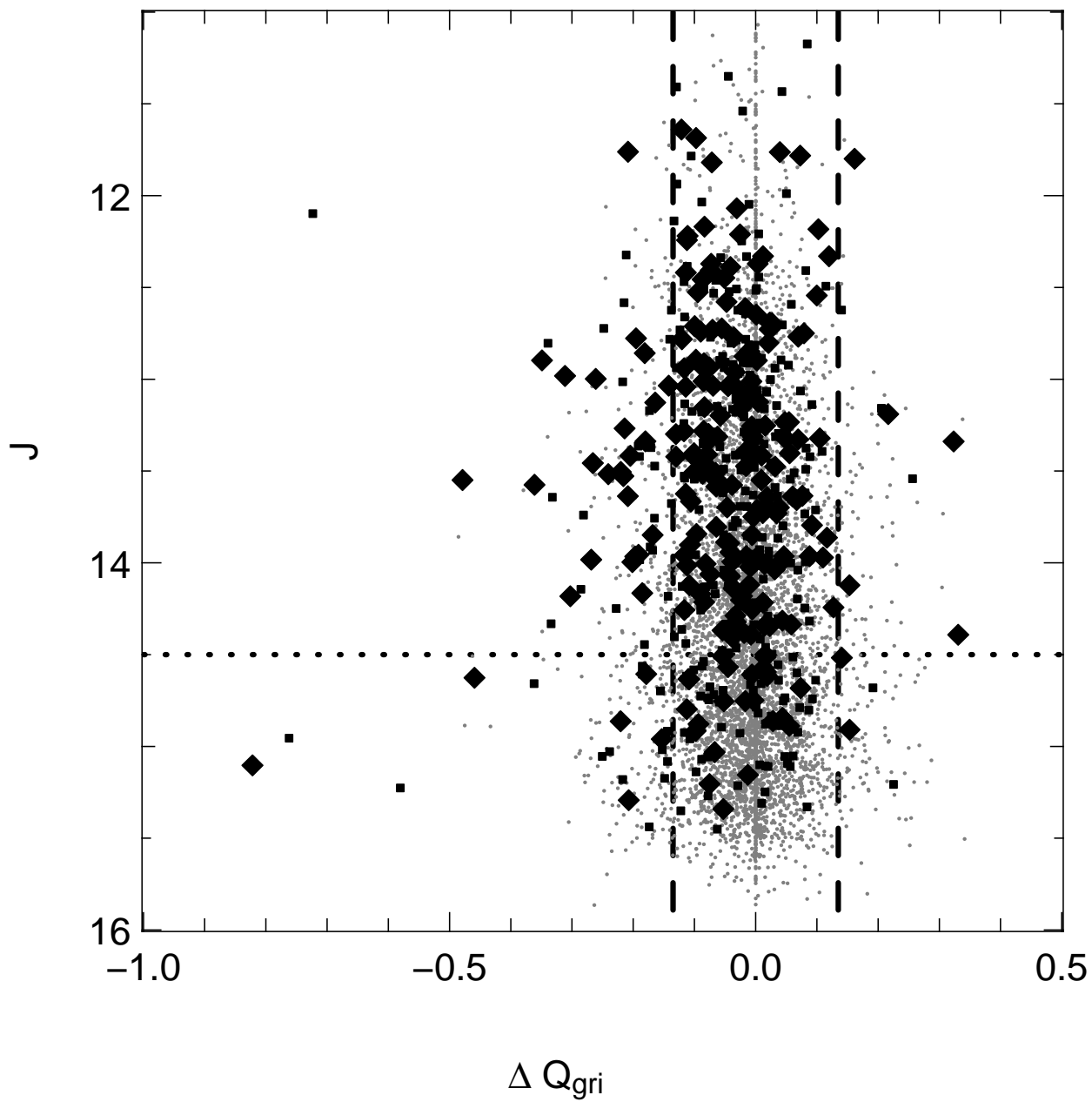}
\caption{{\bf Shift in $Q_{riz}$}.
In this figure we show the shift in $Q_{riz}$ measured between the
bright and faint states
plotted against the
$J$ band magnitude.
The dashed lines are the
3 $\sigma$ limits based on the maximum photometric errors in the survey.
Variable stars are marked following Figure 
\ref{fig-ori_rricmd}. From comparison with Figure \ref{fig-teffqriz}
we see that a change in $Q_{riz}$ of 0.1 corresponds to $\sim$ 100 K
in $T_{eff}$.
\label{fig-ori_delqriz}}
\end{figure}

In Figure \ref{fig-ori_delqriz} we show the 
reddening-invariant $Q_{riz}$ color shift between the faint and 
the bright states defined by the $g$ band brightness against the 2MASS
$J$ magnitude. The observed trend is for $Q_{riz}$ to
decrease when the star is brighter in $g$ which is consistent with the 
presence of veiling in the $r$ band. In this figure the horizontal 
dotted line marks 
the substellar boundary based on the BCAH98 2 Myr isochrone placed at 
m-M = 8.22. We convert the BCAH98 $JHK$ magnitudes from the CIT to the 
2MASS system following \citet{car01}.
The derived effective temperature from $Q_{riz}$ changes by
approximately 100 K for every  0.1 change in the color (see Figure
\ref{fig-teffqriz}). Due to this uncertainty in determination of the
photospheric $Q_{riz}$ color
we will present all color and variability trends against the
2MASS $J$ magnitude which should be relatively unaffected by both
shock emission and disk thermal emission.

\begin{figure}
\plotone{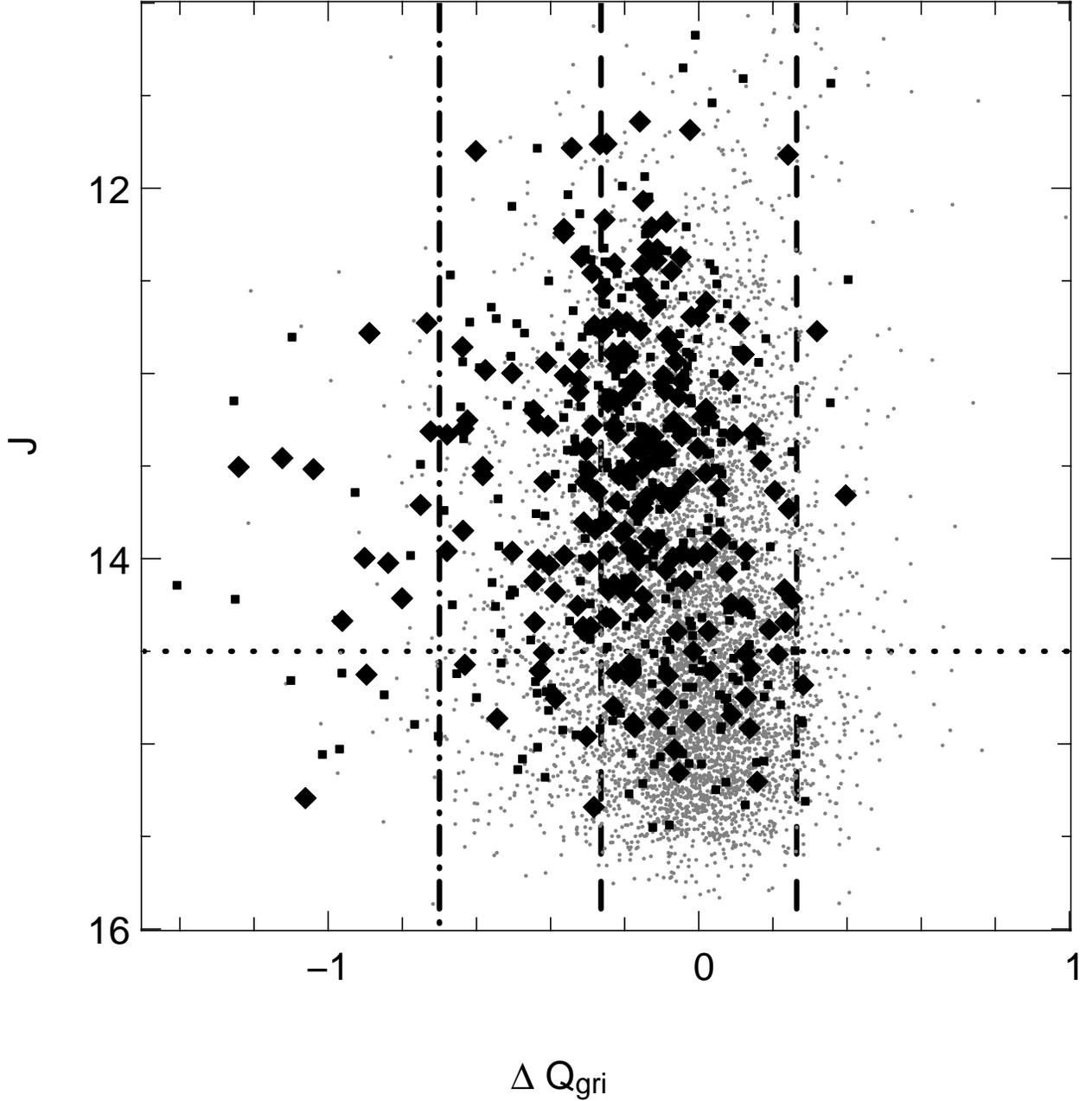}
\caption{{\bf Bright state $\Delta Q_{gri}$}.
The bright state $\Delta Q_{gri}$ is shown here
measured relative to the dwarf locus.
These values correspond to $g$ band veilings
less than 1 (see Figure \ref{fig-model_dobsqxyz}).
The dashed line at $\Delta Q_{gri}$ = -0.7 indicates the
color shift due to surface gravity effects alone
for a pre-main sequence star having
$T_{eff}$ = 2700 K. Variable stars are marked following Figure 
\ref{fig-ori_rricmd}.
The dashed lines centered around $\Delta Q_{gri}$ = 0 are the
3 $\sigma$ limits based on the maximum photometric errors in the survey.
Evidence for weak veiling is found in stars 
fainter the substellar boundary.
\label{fig-ori_invarcc}}
\end{figure}

In Figure \ref{fig-ori_invarcc} we see that the bright state
$\Delta Q_{gri}$ during both faint and bright states slightly 
decreases in the less massive (fainter) objects. However, when the
observed $\Delta Q_{gri}$ are compared against those predicted 
from the cooler ($\sim$ 2700 K) low surface gravity models (dot-dash line)
it is clear that the bulk of veiling signature is probably due to the
shifts in photospheric colors. 
Thus, at the substellar boundary we only see evidence for weak $g$ band 
veiling which consistent with the results of \citet{muz03} who find a 
deficit of continuum veiling at 5500 {\AA} for accreting stars of spectral 
types M5 to M7.

\subsection{Variability}

If the veiling in a specific band $x$ ranges from $r^{low}_x$ to 
$r^{high}_x$
then the observed magnitude change is
\begin{equation}
%\begin{split}
\Delta m_x  = -2.5 log\biggl(\frac{1 + r^{high}_x}{1 + r^{low}_x}\biggr).
%\end{split}
\end{equation} 
This scales montonically, albeit non-linearly, with $r^{high}_x$, if
\begin{equation}
%\begin{split}
\frac{d r^{low}_x}{d r^{high}_x} \le \frac{1 + r^{low}_x}{1 + r^{high}_x}
%\end{split}
\end{equation}
which in the case where $r^{low}_x$ scales with $r^{high}_x$, i.e. 
$r^{low}_x = \alpha r^{high}_x$ requires $\alpha \le 1$, which is implied
by $r^{low}_x \le r^{high}_x$.
In general we then expect the amplitude and RMS of the variability to 
be related to the veiling and hence with the shock flux in that band.

\begin{figure}
\plotone{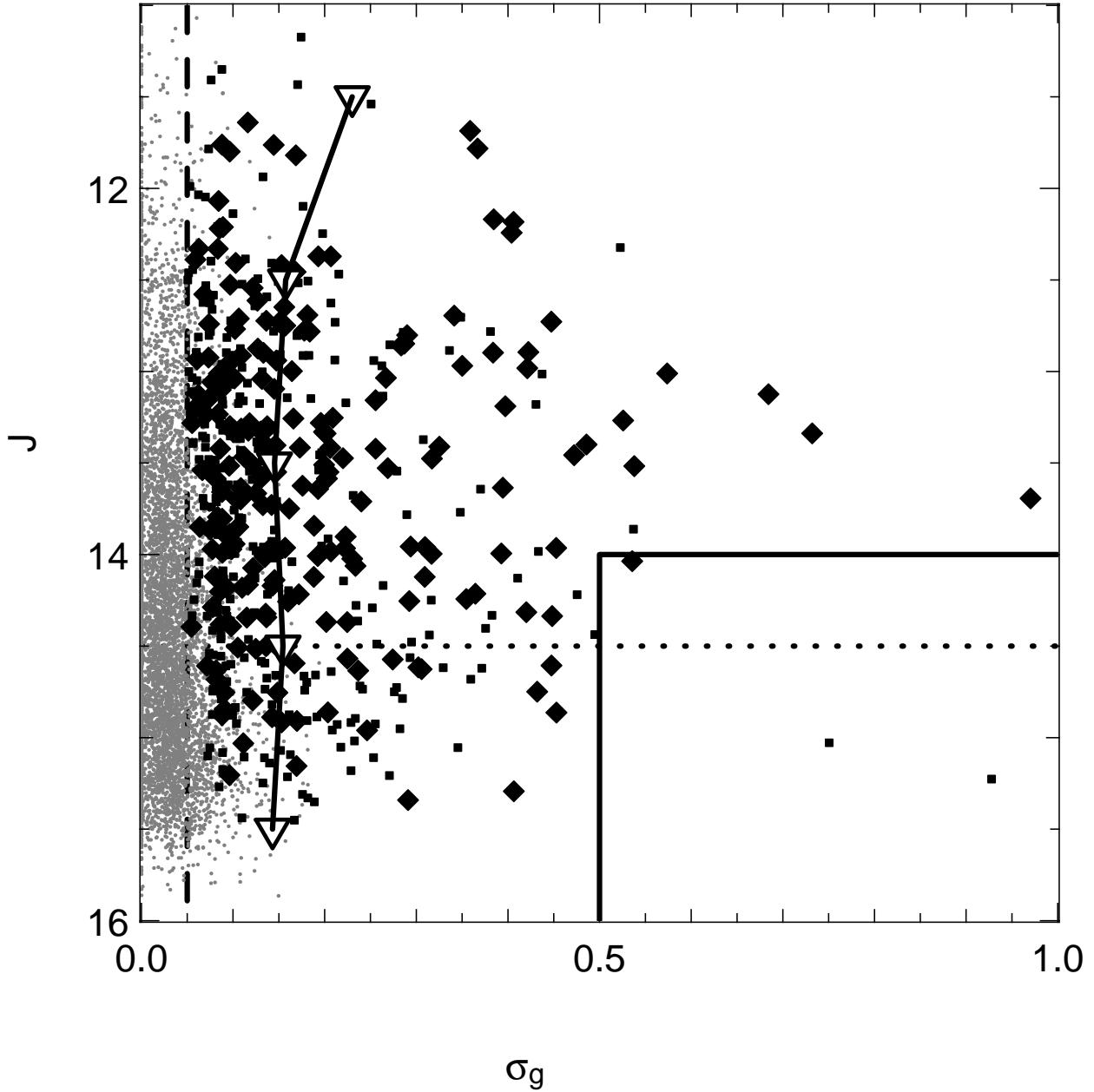}
\caption{{\bf Orion $g$ band variability.} 
The standard deviation of the $g$ band variation is plotted
against $J$ for the Orion pre-main sequence
sample. 
Variable stars are marked following Figure 
\ref{fig-ori_rricmd} above. The median $\sigma g$ values computed
in one-magnitude bins are shown by the connected open triangles.
The dotted line marks the HBL at 2 Myr for
Orion and the dashed line traces the $\sigma_g = 0.05$ threshold
for inclusion in this sample.
The box in the lower right-band corner is defined by $J > 14$ and
$\sigma_g > 0.5$. A lack of faint objects having high-amplitude
variabilities is evident. 
\label{fig-ori_sigg}}
\end{figure}

\begin{figure}
\plotone{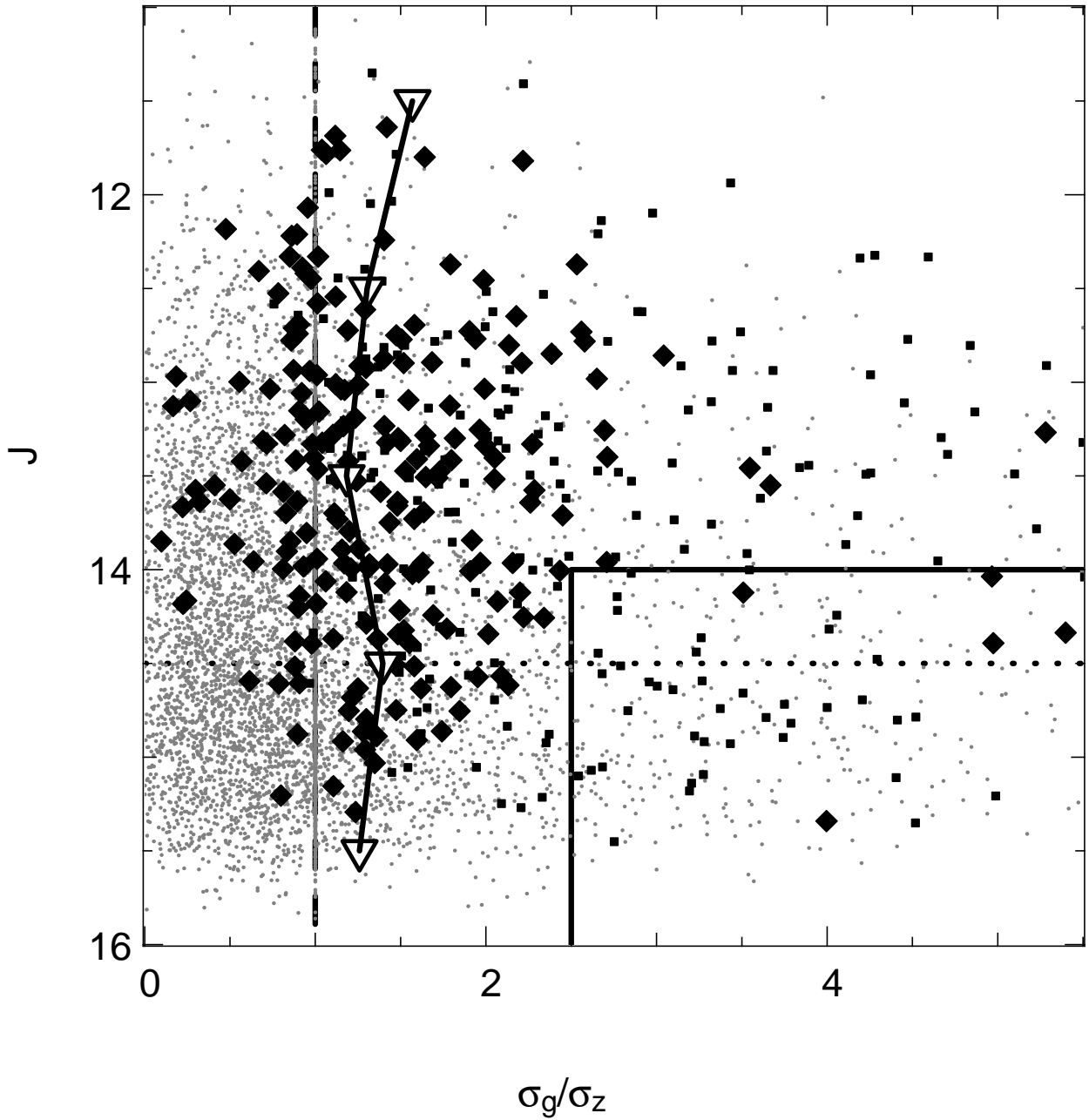}
\caption{{\bf Ratio of $g$ band to $z$ band variability.} 
The $\sigma g$ to $\sigma z$ ratio is plotted
against $J$ for the Orion pre-main sequence
sample.
The median $\sigma g/\sigma_z$ ratios computed
in one-magnitude bins are shown by the connected open triangles.
With the exception of 5 outliers in the lower right
we see that the $g$ band
variation decreases relative to that in $z$ for the fainter 
objects. 
The dotted line marks the HBL at 2 Myr for
Orion. The $\sigma_g = \sigma_z$ locus is shown by the
vertical dashed line.
The box in the lower right-band corner is defined by $J > 14$ and
$\sigma_g/\sigma_z > 3$. A relative lack of objects in this selection
region may reflect a reduction in accretion shock temperature.
\label{fig-ori_siggz}}
\end{figure}

In Figure \ref{fig-ori_sigg} we see that the maximum RMS of the $g$ band 
variations peaks at $J \sim 13$ and that no high-amplitude ($\sigma_g >  
0.5)$ variables are seen below $J = 14$. The triangles mark the median
$\sigma_g$ values computed in for one-magnitude bins spanning
$11 < J < 16$.  
We also see that the maximum
ratio of the $g$ band to $z$ band RMS 
(${\sigma} g$/${\sigma} z$) peaks at the same magnitude and then
decreases in the fainter stars (Figure \ref{fig-ori_siggz}). Only five 
faint ($J > 14$) candidate accreting stars have $\sigma_g/\sigma_z > 3$.\
This trend is perhaps indicative of cooling of the
accretion shock temperature with diminished mass.

For the intermediate mass Classical T Tauris studied by \citet{gul98},
the Balmer jump diminishes as the veiling and shock
bolometric luminosity increase. Conversely, for the weakly accreting
systems expected in very low mass stars and young brown dwarfs the
Balmer jump should be relatively large suggesting that the bulk
of the shock emission will be in the near-ultraviolet rather than
the optical bands. The shock models of \citet{muz03} suggest that shocks
with lower energy flux carried by the accreting column emit more 
at redder wavelengths due to quasi-blackbody emission from the
heated photosphere. The trend we see is consistent with a decrease
in the shock energy flux with decreasing mass.

\subsection{Near-Infrared Excess}

\begin{figure}
\plotone{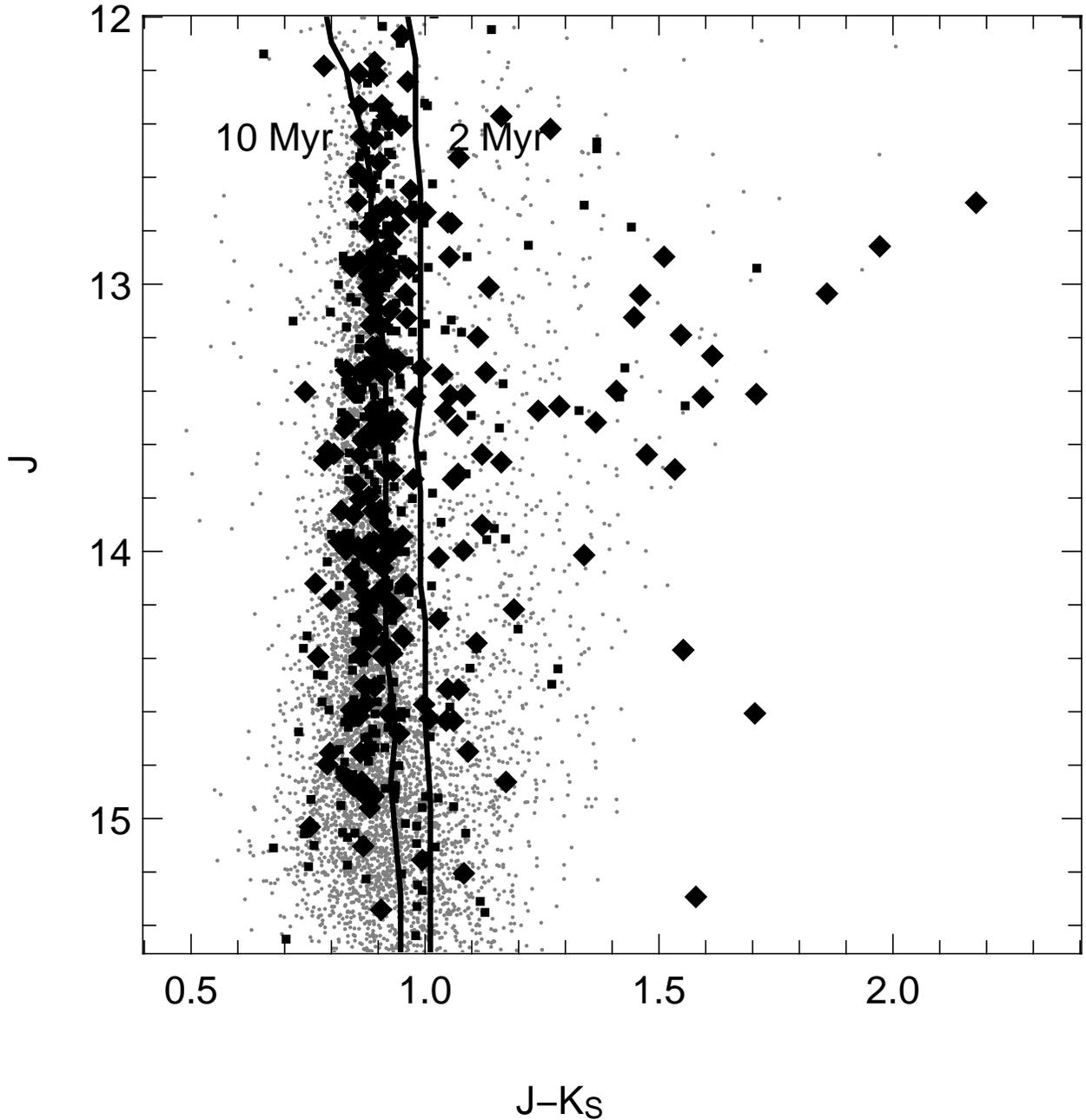}
\caption{{\bf Orion $(J-K_S,J)$ color-magnitude diagram.} 
The 2MASS $J$ magnitude is plotted against 
$J-K_S$ color with the BCAH98 2 Myr and 20 Myr isochrones, converted from
the CIT to the 2MASS system, shown for 
comparison. The dotted line marks the substellar boundary at
this age and the diagonal line indicates the $A_V = 5$ reddening vector.
The objects below the reddening vector are substellar candidates.
Variable stars are marked following Figure 
\ref{fig-ori_rricmd}.
\label{fig-ori_jkj}}
\end{figure}

Near-Infrared ($JHK$) photometry has often been
used to detect the presence of excess emission
primarily in the $K$ band due to thermal
emission from the inner circumstellar disk. \citet{mey97} find a
Classical T Tauri locus in $(J-H,H-K)$ based on the dereddened
colors of K7/M0 intermediate massed stars in Taurus. In Figure
\ref{fig-ori_jkj} we see that the $K_S$ excess indicating either
a disk signatures or significant extinction drops for
objects having $J < 14$. For the 2 Myr BCAH98 isochrone this
corresponds to 0.11 M$_\odot$ and $T_{eff}$ = 3000 K ($\sim$ M5.5).
This lack of signature
in $JHK$ is expected for substellar objects since they are less luminous
and so the inner circumstellar disk will be cooler.
As discussed by \citet{liu03}
this contrast problem between the disk and the stellar photosphere motivates 
searching for young brown dwarf disks
in the $L$ and longer wavelength bands.

\section{Summary}

Analysis of 2MASS and multi-epoch SDSS photometry of low mass 
pre-main sequence objects 
stars within the Orion OB1 assocation shows that:
 
1. $g$ band variations having a standard deviation greater than 0.05
magnitudes persist 
for objects having reddening-invariant colors redder than, and $J$ 
magnitudes fainter than, that expected for the
stellar/substellar boundary. 

2. While identification of $g$ band veiling for spectral types later than
M4 is hindered by our uncertain knowledge of surface gravity effects
on photospheric colors, we see 
evidence for accretion
signatures even in these very low mass objects. The general trend is for the
veiling signature to diminish with decreasing temperature (and mass).

3. No disk signatures based on $J-K$ color excess are seen for 
spectral types later than M4. We infer based on variability and weak
veiling that the accretion process and hence the disks are
present, and that relative contribution to the $K$ band flux is weak.
This is consistent with earlier work on young brown dwarf disks.

We have demonstrated that multi-epoch and multi-band imaging surveys
enable identification and classification of pre-main sequence objects based on 
variability and reddening-invariant colors. While use of the SDSS $g$ band
was mandated by the sensitivity issues and the red-leak in the imager's
filter system, inclusion of a broad-band filter that is shortward
of the Balmer jump such the SDSS $u$ in the next generation of surveys,
i.e. the Large Synoptic Survey Telescope or LSST, will increase the
sensitivity to the veiling emission from the accretion shock for the
later and less massive objects.

We are supplementing this imaging survey with an SDSS spectroscopic
program targeting low mass stars in Orion and the Taurus-Auriga
star formation region. Results from this study in addition to
analysis of the veiling trends presented here and their constraints
on shock luminosities and mass accretion rates will be discussed
in subsequent work.

\acknowledgements

We thank the anonymous referee for comments 
that have greatly improved this paper.
PMM and JAS acknowledge support from LANL Laboratory Directed Research
and Development (LDRD) program 20030486DR. 
Funding for the Sloan Digital Sky Survey (SDSS) has been provided by the Alfred P. Sloan
Foundation, the Participating Institutions, the National Aeronautics and Space Administration, 
the
National Science Foundation, the U.S. Department of Energy, the Japanese Monbukagakusho, 
and the Max Planck Society. 

The SDSS is a joint project of the University of Chicago, Fermilab, the Institute for Advanced Study,
the Japan Participation Group, The Johns Hopkins University, the Korean Scientist Group, Los Alamos National Laboratory, the
Max-Planck-Institute for Astronomy (MPIA), the Max-Planck-Institute for Astrophysics (MPA),
New Mexico State University, the University of Pittsburgh,
Princeton University, University of Portsmouth, 
the United States Naval Observatory, and the
University of Washington.

Facilities: \facility{SDSS}, \facility{2MASS}

%\clearpage

%\clearpage

%\clearpage

%\clearpage

%\clearpage

%\clearpage

%\clearpage

%\clearpage

%\clearpage

%\clearpage

%\clearpage

%\clearpage

%\clearpage

%\clearpage

%\clearpage

%\begin{figure}
%\plotone{f14.eps}
%\caption{{\bf Orion $(J-H,H-K)$ colors.} 
%Here we plot the 2MASS $(J-H,H-K)$ colors.  The near-IR
%color-color diagram includes the Classical T Tauri locus of \citet{mey97},
%the reddening vector of $A_V = 10$ magnitudes ({\it dashed})
%passing through spectral type M6,
%and the M dwarf locus spanning M0 to M9 \citep{leg92}.
%The population extending to the
%top center of the plot is composed of reddened background stars. 
%The stars having near-IR colors consistent with being reddened away from
%the CTTS locus are predominately variable in both the $g$ and
%$z$ bands ({\it circles}).
%\label{fig-ori_jhk}}
%\end{figure}

\clearpage
\oddsidemargin=-1cm
\tabletypesize{\scriptsize}

\begin{deluxetable}{lll}
\tablecaption{Orion OB1 Sub-Associations\label{tbl-oriob1}}
\tablehead{
\colhead{Region} &
\colhead{Distance in pc (m-M)} &
\colhead{Age (Myr)}
}
\startdata
Orion OB1a & 330 (7.59) & 11.4 $\pm$ 1.9 \\
Orion OB1b & 440 (8.22) & 1.7 $\pm$ 1.1 \\
Orion OB1c & 460 (8.31) & 4.6$^{+1.8}_{-2.1}$ \\
Orion OB1d\tablenotemark{a} & 450 (8.27)	& $<$ 1 \\
\enddata
\tablenotetext{a}{The Orion Nebula Cluster is
not included in this study.}
\end{deluxetable}

%\clearpage
\begin{deluxetable}{rrrr}
\tablecaption{SDSS Orion Imaging Runs\label{tbl-runs}\tablenotemark{a}}
\tablehead{
\colhead{Run} &
\colhead{Date (MJD)} &
\colhead{Interval (Days)} &
\colhead{Elapsed (Days)}
}
\startdata
0211 & 51115 & \ldots & \ldots \\
0259 & 51134 & 19 & 19 \\
0273 & 51136 & 2 & 21 \\
0297 & 51139 & 3 & 24 \\
0307 & 51140 & 1 & 25 \\
2955 & 52312 & 1172 & 1197 \\
2960 & 52313 & 1 & 1198 \\
2968 & 52314 & 1 & 1199 \\
4158 & 52912 & 598 & 1787 
\enddata
\tablenotetext{a}{This table lists the SDSS imaging of the Orion equatorial
region giving the run number and the MJD of the observation. The Interval
and Elapsed columns indicate the temporal sampling by giving the
number of days between successive observations and the total number of
days since the first imaging, respectively.}
\end{deluxetable}

%\clearpage

%\begin{deluxetable}{lrcc}
%\tablecaption{Gravity Dependence of Colors at 2700 K\tablenotemark{a}
%\label{tbl-lowg}}
%\tablehead{
%\colhead{Band} &
%\colhead{$\Delta$ B.C.\tablenotemark{b}} &
%\colhead{$\Delta$ (x-y) \tablenotemark{c}} &
%\colhead{$\Delta Q_{xyz}$\tablenotemark{d}} 
%}
%\startdata
%$u$ & -2.782 & \ldots & \ldots \\
%$g$ & -0.489 & $\Delta (u-g) = -2.293$ & \ldots \\
%$r$ & -0.194 & $\Delta (g-r) = -0.295$ & 
%$\Delta Q_{ugr} = -1.999, -2.076$ \\
%$i$ & +0.073 & $\Delta (r-i) = -0.267$ & 
%$\Delta Q_{gri} = +0.199, +0.063$ \\
%$z$ & +0.055 & $\Delta (i-z) = +0.018$ & 
%$\Delta Q_{riz} = -0.285, -0.285$ \\
%$J$ & +0.039 & $\Delta (z-J) = +0.016$ & 
%$\Delta Q_{izJ} = +0.005, +0.008$ \\
%$H$ & +0.122 & $\Delta (J-H) = -0.083$ & 
%$\Delta Q_{zJH} = +0.219, +0.219$ \\
%$K$ & +0.019 & $\Delta (H-K) = +0.103$ & 
%$\Delta Q_{JHK} = -0.114, -0.114$ \\
%\enddata
%\tablenotetext{a}{This table presents the model-based color changes
%between the disk dwarf surface gravity of log(g) = 5.5 and the
%typical pre-main sequence surface gravity of log(g) = 3.5.}
%\tablenotetext{b}{Change in bolometric correction.}
%\tablenotetext{c}{Change in color.
%A negative value means that the color becomes
%bluer when the surface gravity is lowered.}
%\tablenotetext{d}{Change in reddening-invariant index. The two
%entries are for $R_V$ = 3.1 and $R_V$ = 5.5, respectively.}
%\end{deluxetable}

%\clearpage

\begin{deluxetable}{llcc}
\tablecaption{Limiting Magnitudes and Photometric Errors\label{tbl-limits}\tablenotemark{a}}
\tablehead{
\colhead{Band} &
\colhead{Survey} &
\colhead{Faint Limit} &
\colhead{Maximum Error}
}
\startdata
$g$ & SDSS & 21.5 & 0.05 \\
$r$ & SDSS & 20.0 & 0.03 \\
$i$ & SDSS & 18.0 & 0.02 \\
$z$ & SDSS & 17.0 & 0.02 \\
$J$ & 2MASS & 15.5 & 0.05 \\
$H$ & 2MASS & 14.5 & 0.05 \\
$K_S$ & 2MASS & 14.5 & 0.10 
\enddata
\tablenotetext{a}{These are the faint magnitude limits and 
maximum photometric errors observed when the selection criteria
described in the text are applied to the SDSS and 2MASS data
acquired in the Orion equatorial region.}
\end{deluxetable}

%\clearpage

\begin{deluxetable}{rr}
\tablecaption{Scaling Between Actual and Observed $\sigma$\label{tbl-varthres}
\tablenotemark{a}}
\tablehead{
\colhead{N Epoch} &
\colhead{Factor (99\%)}
}
\startdata
2 & 1.748 \\
3 & 1.344 \\
4 & 1.226 \\
5 & 1.155 \\
10 & 1.072 \\
15 & 1.040 \\
20 & 1.033 \\
25 & 1.026 \\ 
30 & 1.021
\enddata
\tablenotetext{a}{This table was created, as described in the
text, by a series of Monte Carlo simulations to determine the maximum
standard deviation ($\sigma$) in a random sequence given an intrinsic 
$\sigma$ of the parent population
and a limited number of observations (N Epoch). The scaling factors
reflect the maximum $\sigma$ seen in 99\% of the simulated data sets at
each value of N Epoch. In the limit of large N Epoch the scaling factors
tend towards unity.}
\end{deluxetable}

\end{document}